\documentclass[apj,iop,numberedappendix,appendixfloats]{emulateapj}

\usepackage{amsmath}
\usepackage{multirow}

\bibpunct[,]{(}{)}{;}{a}{}{,}

\newcommand{\Omegam}{\ensuremath{\Omega_{\mathrm{m}}}}
\newcommand{\OmegaL}{\ensuremath{\Omega_{\Lambda}}}

\newcommand{\hmpc}{\; h^{-1}\mathrm{Mpc}}

\newcommand{\hMsun}{\; h^{-1}M_{\odot}}
\newcommand{\kms}{{\; {\rm km}\,{\rm s}^{-1}}}
\newcommand{\Hunits}{\kms {\mathrm{Mpc}}^{-1}}
\newcommand{\hden}{\; h^{3}{\mathrm{Mpc}}^{-3}}

\newcommand{\hvolg}{\; h^{-3}{\mathrm{Gpc}}^{3}}

\newcommand{\mpart}{m_\mathrm{part}}

\newcommand{\lcdm}{\ensuremath{\Lambda\mathrm{CDM}}}
\newcommand{\Lstar}{\ensuremath{L_{\ast}}}

\newcommand{\Dchisq}{\ensuremath{\Delta\chi^2}}

\newcommand{\ntropy}{{\it N}tropy}
\newcommand{\nbody}{$N$-body}
\newcommand{\npoint}{$n$-point}
\newcommand{\pimax}{\pi_\mathrm{max}}

\newcommand{\band}[2]{\ensuremath{^{#1}\!{#2}}}
\newcommand{\Mr}{M_{\band{0.1}{r}}}

\newcommand{\plotwidth}{0.85\linewidth}
\newcommand{\hplotwidth}{0.425\linewidth}

\begin{document}

\title{Three-Point Correlation Functions of SDSS Galaxies: Constraining Galaxy-Mass Bias}
\submitted{Submitted to ApJ: 14 December 2010}

\author{
  Cameron~K.~McBride\altaffilmark{1,2},
  Andrew~J.~Connolly\altaffilmark{3},
  Jeffrey~P.~Gardner\altaffilmark{4},
  Ryan~Scranton\altaffilmark{5},
  Rom\'an~Scoccimarro\altaffilmark{6},
  Andreas~A.~Berlind\altaffilmark{2},
  Felipe~Mar\'in\altaffilmark{7,8}
  Donald~P.~Schneider\altaffilmark{9}
}
\email{cameron.mcbride@vanderbilt.edu}


\altaffiltext{1}{Department of Physics and Astronomy, University of Pittsburgh, Pittsburgh, PA 15260} 
\altaffiltext{2}{Department of Physics and Astronomy, Vanderbilt University, Nashville, TN 37235}
\altaffiltext{3}{Department of Astronomy, University of Washington, Seattle, WA 98195-1580}
\altaffiltext{4}{Department of Physics, University of Washington, Seattle, WA 98195-1560}
\altaffiltext{5}{Department of Physics, University of California, Davis, CA 95616}
\altaffiltext{6}{Center for Cosmology and Particle Physics, New York University, New York, NY 10003, USA}
\altaffiltext{7}{Department of Astronomy \& Astrophysics, Kavli Institute for Cosmological Physics, The University of Chicago, Chicago, IL 60637 USA}
\altaffiltext{8}{Centre for Astrophysics \& Supercomputing, Swinburne University of Technology, Hawthorn, VIC 3122, Australia}
\altaffiltext{9}{Department of Astronomy \& Astrophysics, Pennsylvania State University, University Park, PA 16802}


\begin{abstract}

We constrain the linear and quadratic bias parameters from the 
\emph{configuration} dependence of the three-point correlation function (3PCF) 
in both redshift and projected space, utilizing measurements of spectroscopic 
galaxies in the Sloan Digital Sky Survey (SDSS) Main Galaxy Sample. 
%
%
We show that bright galaxies ($M_r < -21.5$) are biased tracers of mass, 
measured at a significance of $4.5\sigma$ in redshift space and 
$2.5\sigma$ in projected space by using a thorough error analysis in the 
quasi-linear regime ($9-27\; \hmpc$).
Measurements on a fainter galaxy sample are consistent with an unbiased model.  
We demonstrate that a linear bias model appears sufficient to explain the 
galaxy-mass bias of our samples, although a model using both linear and 
quadratic terms results in a better fit.
In contrast, the bias values obtained from the linear model appear in better 
agreement with the data by inspection of the relative bias, and yield
implied values of $\sigma_8$ that are more consistent with current constraints.
We investigate the covariance of the 3PCF, which itself is a measurement of 
galaxy clustering.  We assess the accuracy of our error estimates
by comparing results from mock galaxy catalogs to jackknife re-sampling methods. 
We identify significant differences in the structure of the covariance.  However, 
the impact of these discrepancies appears to be mitigated by an eigenmode analysis 
that can account for the noisy, unresolved modes.  Our results demonstrate that
using this technique is sufficient to remove potential systematics 
even when using less-than-ideal methods to estimate errors. 

\end{abstract}

\section{Introduction}
\label{s:intro}

Studying the statistical properties of the galaxy distribution allows one to probe 
the structure of overdense regions today, learning about galaxy formation and cosmology.  
We observe significant clumping in this large-scale structure (LSS), which is commonly 
characterized by a series of \npoint\ correlation functions \citep[reviewed in][]{peebles:80}.  
Observational evidence is in line with predictions of a dark-energy dominated cold dark 
matter ($\lcdm$) model \citep{komatsu:09,sanchez:09,reid:10}.  
However, there is a large conceptual hurdle between following the evolution of mass densities 
in gravitational collapse \citep[e.g. ][]{lss_review} and that realized by galaxy 
positions.  A priori, there is little reason to believe a one-to-one correspondence exists 
between mass overdensities and galaxy positions; complex galaxy formation processes such 
as merging and feedback should have significant contributions.  
For example recent results from the Sloan Digital Sky Survey \citep[SDSS; ][]{york:00} in 
\citet{zehavi:05,zehavi:10} show clustering varies with galaxy luminosity and color.  This 
discrepancy between the observed ``light'' in galaxies relative to the predicted ``mass'' 
clustering is often described as \emph{galaxy-mass bias}.

The parameterization of galaxy-mass bias enables a two-pronged approach to 
probe both cosmology and galaxy formation.  On one side, we map the clustering of 
galaxies to that of the underlying mass distribution allowing us to understand and 
constrain cosmology.  Alternatively, the parameterization of the bias itself encodes 
useful information concerning galaxy formation processes.
This approach distills observational data from hundreds of thousands of galaxies 
available in modern surveys, such as 
the the two-degree field galaxy redshift survey \citep[2dFGRS;][]{2dFGRS} 
and the SDSS into a significantly smaller and more manageable form. 

Most observational evidence exploits the two-point correlation function (2PCF), 
the first in the series of \npoint\ functions (or equivalently, the power spectrum 
in Fourier space).  However, the 2PCF represents only a portion of the available information. 
Measurements of higher order moments, such as the three-point correlation function 
(3PCF), allow a more complete picture of the galaxy distribution. The statistical 
strength of higher order information might rival that of two-point statistics \citep{sefusatti:05}, 
as well as break model degeneracies describing cosmology and galaxy bias 
\citep{zheng:07,kulkarni:07}.  

Previous analyses have estimated the 3PCF from modern galaxy redshift surveys, including 
work on the the 2dFGRS \citep{jing:04,wang:04,gaztanaga:05} and results from SDSS 
data \citep{kayo:04,nichol:06,kulkarni:07,gaztanaga:09,marin:11}.  Related higher order 
statistics have also been measured for these datasets \citep{verde:02,pan:05,hikage:05,nishimichi:07}.

This work is the second of two papers analyzing the reduced 3PCF on SDSS galaxy samples.
The first paper \citep{mcbride:10} focused on the details of the measurements we 
analyze here, as well as clustering differences due to galaxy luminosity and color.  
This paper utilizes the configuration dependence to constrain non-linear galaxy-mass 
bias parameters in the local bias model \citep{fry:93}, and the properties of the 
errors necessary for quantitative analyses.  

The local bias model is a simple approach to characterize galaxy-mass bias.  
Alternative descriptions exist based on the halo model \citep[reviewed in][]{cooray:02}, 
which form phenomenological models with a wider range of parameters. Two well used 
formulations include the halo occupation distribution \citep[HOD;][]{berlind:02} 
and the conditional luminosity function \citep[CLF;][]{yang:03,vdB:03}. 
There are formulations for the 3PCF; however, the accuracy of the model predictions is 
not as well determined as the 2PCF when compared with data \citep[see][]{takada:03,wang:04,fosalba:05}.  
A significant advantage of a HOD modeling is the ability to use well determined 
measurements of the small scales for constraints (the non-linear regime in gravitational 
perturbation theory).  Understanding the projected 3PCF, $Q_{proj}$, a major component 
of this work, provides a critical link to obtain reliable measurements at these 
smaller scales from observational galaxy samples.  

However, by using this simple prescription for galaxy-mass bias, we investigate the effects 
of binning and covariance resolution in a quantitative analysis with a clear and simple model 
where the implications for bias and cosmology are better studied for higher order moments.
An important part of our analysis is comparing results from the projected 3PCF with the more 
commonly used redshift space measurements. 

This paper is organized as follows. 
We discuss the SDSS data, simulations, and mock galaxy catalogs in \S\ref{s:data}.  
We review the theory and methods of our analysis in \S\ref{s:methods}.  
We constrain the non-linear galaxy mass bias parameters in \S\ref{s:bias}. 
In \S\ref{s:ev}, we investigate clustering properties contained in the eigenvectors of the 
3PCF covariance matrix.  
We perform a detailed examination of the quality of error estimation in \S\ref{s:errors}. 
We discuss our results and compare to related analyses in \S\ref{s:disc}.
Finally, we review our main conclusions in \S\ref{s:summary}.
Unless otherwise specified, we assume a flat $\lcdm$ cosmology where 
$\Omegam = 0.3$, $\OmegaL = 0.7 $, and $H_o = h \, 100 \Hunits$, 
used to convert redshift to physical distances.

\section{Data}
\label{s:data}

\subsection{SDSS Galaxy Samples}
\label{s:sdss}

\begin{table*}
  \centering
  \begin{tabular}{lccccc}
    \hline
    \multicolumn{6}{c}{\bfseries Specifics of SDSS galaxy samples}  \\
    \multirow{2}{*}{Sample} & Absolute  & \multirow{2}{*}{Redshift} & Volume   & Number of & Density \\ 
		            & Magnitude &                           & $\hvolg$ & Galaxies  & $10^{-3} \hden $ \\ 
    \hline
    \hline
    BRIGHT        &         $M_{r} < -21.5$ & $0.010 $ - $0.210 $ & $0.1390$ & $ 37,875$ & $0.272$  \\
    LSTAR         & $-21.5 < M_{r} < -20.5$ & $0.053 $ - $0.138 $ & $0.0391$ & $106,823$ & $2.732$  \\
    \hline
  \end{tabular}
  \caption[Volume-limited galaxy catalogs from the SDSS DR6]{ 
    The magnitude range, redshift limits, volume, total number of galaxies, and completeness corrected number 
    density are shown for the galaxy samples constructed from the SDSS DR6 spectroscopic catalog.  
    We selected these samples by cuts in redshift and corrected (K-correction and passive evolution) absolute $r$-band 
    magnitude to create volume-limited selections.  See details in \citet{mcbride:10}.
  } \label{t:gal_samples}
\end{table*}

The SDSS has revolutionized many fields in astronomy, obtaining images and spectra 
covering nearly a quarter of the sky by utilizing a dedicated 2.5 meter telescope 
at Apache Point Observatory in New Mexico \citep{gunn:98,gunn:06,york:00,stoughton:02}. 

Our galaxy samples and details of the measurements are fully described in a companion 
paper \citep{mcbride:10}.
Briefly, we use galaxy data with spectroscopically determined redshifts, defined as the Main 
galaxy sample \citep{strauss:02}.  We conduct our analysis of clustering measurements using 
galaxies from DR6 \citep{sdss_dr6}, and define samples from a refined parent catalog: the 
New York University Value-Added Galaxy Catalog \citep[NYU-VAGC;][]{vagc}.  
We analyze two samples: a BRIGHT sample where $M_r < -21.5$ and LSTAR with $-21.5 < M_r < -20.5$.  
We do not analyze the FAINT sample presented in the companion paper \citep{mcbride:10}, 
as the errors suffer from small volume effects.  
We tabulate properties, such as the redshift range, number of objects, volume and completeness 
corrected number density in Table~\ref{t:gal_samples}.

Our absolute $r$-band magnitudes use the NYU-VAGC convention defined to represent values at 
$z=0.1$ \citep[see details in][]{vagc}.  We note these as $M_r$ for simplicity, which 
refer to $\Mr - 5 \log h$.  Radial distances and absolute magnitudes are calculated using a 
flat \lcdm\ cosmology with $ \Omegam = 0.3 $ and $ \OmegaL = 1 - \Omegam $.

\subsection{Hubble Volume Simulation}
\label{ss:hvs}

To estimate the clustering of mass in the late time \lcdm\ cosmology, we analyze 
cosmological \nbody\ simulations.  We use the \emph{Hubble Volume} (HV) 
simulations \citep{colberg:00,evrard:02} that were completed by the the Virgo Consortium.  
We utilize the lightcone output with \lcdm\ cosmology: 
($\Omegam = 0.3$, $\OmegaL = 0.7$, $h = 0.7$, $\sigma_8 = 0.9$), where 
$H_o = h 100 \Hunits$.  The HV simulation consists of $1000^3$ particles in a box of 
$(3000 \hmpc)^3$ volume, resulting in a particle mass of 
$\mpart = 2.2 \times 10^{12} \hMsun$.  The particles start from an 
initial redshift of $z_{init} = 35$, and are evolved to the current time using a 
Plummer softened gravitational potential with a softening length of $0.1 \hmpc$.

We use the same simulation output as presented in our companion paper \citep{mcbride:10}.
Here, we briefly review our postprocessing of the simulation data for completeness. We 
include redshift distortions in the mass field by distorting the position according to 
the peculiar velocity of the dark matter particle.  We trim particles to match the 
identical volume of the corresponding SDSS samples, including the non-trivial angular 
geometry of SDSS data.  Finally, we randomly downsample the number of dark matter 
particles to make the computational time of the analysis more manageable 
\citep[discussed further in ][]{mcbride:10}.

\subsection{Mock Galaxy Catalogs}
\label{ss:mock_desc}

We analyze mock galaxy catalogs created to match some of the SDSS galaxy data. 
These mock catalogs were constructed from 49 independent \nbody\ simulations, initiated 
with different random phases and evolved from a single cosmology: 
($\Omega_m = 0.27, \; \Omega_\Lambda = 0.73, \; h=0.72, \; \sigma_8 = 0.9$).
While these differ slightly from our assumed cosmology for the data, we expect the 
differences to be minor, with no significant implications on our analysis. 
Each of the 49 realizations had randomized phases where the initial conditions 
were generated from 2nd order Lagrangian perturbation theory 
\citep{scoccimarro:98,crocce:06}.  
These simulations each consist of $640^3$ particles that we evolved using Gadget2 
\citep{gadget2} from an initial redshift of $z_i=49$ to the present epoch.  The box 
side-length of $1280 \; \hmpc$ contained enough volume to exactly match the brightest 
galaxy sample after applying the SDSS geometry.  These simulations have been used in 
various other studies \citep[e.g.][]{tinker:08,manera:10}.

The galaxy mocks were created by populating dark matter halos with galaxies by 
applying the HOD model in \citet{tinker:05} with parameter values defined  
to represent $M_r < -21.5$ and $\sigma_8=0.9$. 
The halos were identified using a friends-of-friends algorithm \citep{fof} applying a 
linking length of $b = 0.2$ in units of the mean interparticle separation. The least 
massive halos contained $33$ particles, a minimum mass capable of hosting the faintest 
galaxies in the BRIGHT galaxy sample.  
Given the mass resolution of these simulations, less massive halos necessary to host 
galaxies in the LSTAR galaxy sample could not be identified.  Therefore, we could only 
obtain reliable mock galaxy catalogs corresponding to the BRIGHT sample.

\section{Theory \& Methods}
\label{s:methods}

The \npoint\ correlation functions remain a standard description of the 
complexity seen in large-scale structure \citep[LSS;][]{peebles:80}.  In terms of 
the fractional overdensity ($\delta$) about the mean density ($\bar{\rho}$), 
\begin{equation} \label{eq:overdensity}
  \delta(\vec{x}) = \frac{\rho(\vec{x})}{\bar{\rho}} - 1 \; , 
\end{equation}
we characterize the two-point correlation function (2PCF) 
and three-point correlation function (3PCF) as: 
\begin{equation} \label{eq:2pcf_delta}
  \xi(r_{12}) = \langle \delta(\vec{x}_1) \delta(\vec{x}_2) \rangle \; .
\end{equation}
\begin{equation} \label{eq:3pcf_delta}
  \zeta(r_{12}, r_{23}, r_{31}) = \langle \delta(\vec{x}_1) \delta(\vec{x}_2) \delta(\vec{x}_3) \rangle \; .
\end{equation}
We make the standard assumption of a homogeneous and isotropic distribution, and 
report clustering amplitudes dependent on the magnitude of the separation vector, 
e.g. $r_{12} = |\vec{x}_1 - \vec{x}_2|$.

Motivated by the \emph{hierarchical ansatz} \citep{peebles:80} and gravitational 
perturbation theory \citep{lss_review} we use the \emph{reduced} 3PCF:
\begin{equation} \label{eq:Q}
  Q(r_{12}, r_{23}, r_{31}) = \frac{  \zeta(r_{12}, r_{23}, r_{31}) }
				   {\xi_{12} \xi_{23} + \xi_{23} \xi_{31} + \xi_{31} \xi_{12} }\; .
\end{equation}
This ``ratio statistic'' remains close to unity at all scales, and to leading order is 
insensitive to both time evolution and cosmology \citep[reviewed in][]{lss_review}. 

Redshift distortions impact measurements of clustering by altering the line-of-sight 
radial distant estimate, as we are unable to distinguish the galaxy's peculiar velocity 
from the Hubble flow \citep[reviewed in][]{hamilton:98}. We refer to the theoretical 
non-distorted distances as \emph{real} space, commonly denoted with $r$.  Distances that 
include the redshift distortion (e.g. observational distances) are in \emph{redshift 
space}, denoted with $s$.  We decompose the redshift space distance into line-of-sight 
($\pi$) and projected separation ($r_p$) such that $ s = (\pi^2 + r_p^2)^{1/2} $.  
With this separation, the anisotropic distortion is primarily contained in the 
$\pi$ coordinate.

We minimize the impact of redshift distortions by estimating the correlation function 
binned in both $r_p$ and $\pi$ and integrate along the line-of-sight resulting in 
the projected correlation function \citep{davis:83}: 
\begin{equation} \label{eq:wp}
  w_p(r_p) = 2 \int_{0}^{\pimax} \xi(r_p, \pi) \mathrm{d} \pi \; .
\end{equation}
The projected 3PCF and its reduced form have analogous definitions: 
\begin{multline}   \label{eq:zeta_proj} 
  \zeta_{proj}(r_{p12}, r_{p23}, r_{p31}) = \\ 
    \iint \zeta(r_{p12}, r_{p23}, r_{p31}, \pi_{12}, \pi_{23}) \mathrm{d} \pi_{12} \mathrm{d} \pi_{23}  \; , 
\end{multline}
\begin{multline}   \label{eq:Qproj}
  Q_{proj}(r_{p12}, r_{p23}, r_{p31}) = \\
    \frac{ \zeta_{proj}(r_{p12}, r_{p23}, r_{p31}) }
	 { w_{p12} w_{p23} + w_{p23} w_{p31} + w_{p31} w_{p12} } \; .
\end{multline}
The measurements we analyze set $\pimax = 20 \hmpc$. We find this sufficiently deep to 
recover correlated structure to minimize redshift distortions, but not overly expensive 
to calculate \citep[see detailed discussion in appendix of ][]{mcbride:10}.

The full 3PCF is a function of three variables that characterize both the size and 
shape of triplets.  We parameterize the 3PCF by ($r_1$, $r_2$, $\theta$), where 
$r_1$ and $r_2$ represent two sides of a triangle (simplified notation from $r_{12}$ and 
$r_{23}$), and $\theta$ defines the opening angle between these sides.  
However, our measurements are estimated in bins defined by ($r_{12}, r_{23}, r_{31}$).
We convert $r_{31}$ to $\theta$ using the cosine rule \citep[as detailed in][]{mcbride:10}.
The 3PCF remains sensitive to the exact choice of binning scheme, which can mask or 
distort the expected signal \citep{GS05,marin:08,mcbride:10}.  We choose a bin-width 
as a fraction, $f$, of the measured scale, $r$, such that $\Delta_r = f \times r$ and 
a bin at $r$ represents $(r - \frac{\Delta_r}{2}, r + \frac{\Delta_r}{2})$.  

We always use the reduced 3PCF as a function of three variables, $Q(r_1, r_2, \theta)$, 
but we simply our notation by sometimes referring to it as $Q(\theta)$ or even $Q$.  
If the amplitude of $Q(\theta)$ varies significantly with $\theta$, we refer to 
this as \emph{strong} configuration dependence, in contrast to little or no variation 
for a \emph{weak} configuration dependence.  We define the scale of triangles by $r_1$, 
and choose configurations such that $r_2 = 2 r_1$.  This results in $r_3$ varying in size 
from $r_3 = r_2 - r_1$ when $\theta = 0$ to $r_3 = r_2 + r_1$ when $\theta = \pi$.  

\subsection{Galaxy-Mass Bias}
\label{ss:bias_intro}

We can consider galaxies to be a \emph{biased} realization of the \lcdm\ mass field.
In the local bias model \citep{fry:93}, the galaxy over-density, $\delta_g$, can be 
connected to the mass over-density, $\delta_m$, by a non-linear Taylor series 
expansion:
\begin{equation} \label{eq:bias_mod}
  \delta_g = \sum_k \frac{ b_k }{k!} \delta^k_{m} 
             \approx b_1 \delta_{m} + \frac{b_2}{2} \delta^2_{m} \; .
\end{equation}
This relation describes the mapping between galaxy and mass by simple scalar values, 
to second order: the linear ($b_1$) and quadratic ($b_2$) bias.  

With measurements on galaxy $n$-point correlation functions, the clustering of galaxies 
is linked to mass clustering via the bias parameters.  The 2PCF can be used to 
constrain the linear bias by equating the correlation function between galaxies, $\xi_g$, 
to that of dark matter, $\xi_m$, such that
\begin{equation} \label{eq:bias2pt}
  \xi_g(r) = b_1^2 \, \xi_{m}(r) \; .
\end{equation}
The 3PCF is the lowest order correlation function that shows leading order sensitivity to 
the quadratic bias term. The analog to \eqref{eq:bias2pt} for the connected 3PCF is written
\begin{align} \label{eq:bias3pt}
  \zeta_g(r_{12},r_{23},r_{31}) = & \: b_1^3 \zeta_m(r_{12},r_{23},r_{31}) \: + \\
    & \: b_1^2 b_2 \left[ \xi_{12} \xi_{23} + \xi_{12} \xi_{31} + \xi_{31} \xi_{23} \right] \; , \nonumber 
\end{align}
where $\xi_{12} = \xi_m(r_{12})$, etc.  This simplifies for the reduced 3PCF where we denote the 
bias parameters as $B = b_1$ and $C = b_2 / b_1$:  
\begin{equation} \label{eq:biasQ}
  Q_g(r_{12},r_{23},\theta) = \frac{1}{B} \big[ \; Q_{m}(r_{12},r_{23},\theta) + C \; \big] \; .
\end{equation}
We have changed notation slightly in \eqref{eq:biasQ}, replacing $r_{31}$ with 
$\theta$, the opening angle between the two sides $r_{12}$ and $r_{23}$, as we discussed 
above. 

A multiplicative factor such as $B$ can dampen ($B>1$) or enhance ($B<1$) the 
configuration dependence of $Q_m$ as seen from the galaxy distribution, whereas the 
value of $C$ will produce an offset.  We see that $B$ and $C$ are partially degenerate 
in this model. If $Q(\theta)$ shows no configuration dependence, two parameters are used 
to describe a shift in amplitude.  However, this degeneracy can be removed when the 
3PCF exhibits a shape dependence \citep{fry:94}.  Even with the degeneracy broken, the 
values of $B$ and $C$ could show a strong correlation.
%

\subsection{Estimating the Covariance Matrix}
\label{ss:covar_est}

We measure the correlation between measurements by empirically calculating the covariance 
matrix.  Given a number of realizations, $N$, a fractional error on $Q$ can be written as
\begin{equation}\label{eq:del}
  \Delta_i^k = \frac{ Q_i^k - \bar{Q}_i }{ \sigma_i } \; ,
\end{equation}
for each realization ($k$) and bin ($i$) given a mean value ($\bar{Q}_i$) and variance 
($\sigma_i^2$) for each bin over all realizations.  We use $Q$ as a general placeholder 
for any measured statistic
(2PCF, 3PCF, etc).  

We construct the normalized covariance matrix using the standard unbiased estimator:
\begin{equation}\label{eq:cov}
  \mathcal{C}_{ij} = \frac{1}{N - 1} \sum_{k=1}^{N}\Delta_i^k \Delta_j^k \; .
\end{equation}
Equation~\ref{eq:cov} assumes that each realization is independent.  In practice, a 
number of mock galaxy catalogs can be used to make this a tractable approach.  If mock 
catalogs appropriate to the galaxy sample are not available, a covariance matrix can be 
estimated from the data itself, such as the commonly employed jackknife re-sampling 
\citep{lupton:01}.  Since jackknife samples are not independent realizations, 
we compute the covariance by:
\begin{equation}\label{eq:cov_jack}
  \mathcal{C}^{(jack)}_{ij} = \frac{(N - 1)^2}{N} \mathcal{C}_{ij} 
			    = \frac{N - 1}{N} \sum_{k=1}^{N}\Delta_i^k \Delta_j^k \; , 
\end{equation}
where $\mathcal{C}_{ij}$ denotes the typical unbiased estimator of the covariance when 
computed on $N$ jackknife samples.  

\subsection{Eigenmode Analysis}
\label{ss:ema}

We constrain galaxy-mass bias parameters using the information in the full 
covariance matrix.  We utilize an \emph{eigenmode} analysis \citep{scoccimarro:00}, 
an equivalent method to a principal component analysis (PCA) on the measurement 
covariance matrix. This method was tested in detail for the galaxy-mass bias of the 
3PCF using simulated data in \citet{GS05}.

The basic idea is to isolate the primary contributing eigenmodes of the reduced 3PCF based 
on the structure of the normalized covariance matrix.  This allows one to trim unresolved 
modes and perform a fit in a basis which minimizes the non-Gaussianity of the residuals.  
To summarize, the covariance matrix can be cast in terms of a singular value decomposition 
(SVD), 
\begin{equation}\label{eq:svd}
  \bm{\mathcal{C}} = \bm{ U \; \Sigma \; } \bm{V}^T \quad ; \quad \Sigma_{ij} = \lambda_{i}^2 \delta_{ij} \; .
\end{equation}
where $\delta_{ij}$ is the Kronecker delta function making $\bm \Sigma$ a diagonal matrix 
containing the singular values, $\lambda_i^2$.  The matrices $\bm{U}$ and $\bm{V}$ are
orthogonal rotations to diagonalize the covariance into $\bm{\Sigma}$ where $\bm{V}^T$ 
denotes the transpose of $\bm{V}$.

Applying the SVD to the covariance matrix yields a rotation into a basis where the 
eigenmodes are uncorrelated (i.e. the covariance matrix becomes diagonal).  The 
resulting rotation matrix can be directly applied to our signal forming the 
\emph{$Q$-eigenmodes},
\begin{equation}\label{eq:q-eigenmodes}
  \widehat{Q}_i = \sum_j U_{ij} \frac{ Q_j } { \sigma_j } \; .
\end{equation}
The singular values provide a weight on the importance of each eigenvector.  
Specifically, a multiplicative factor of $1/\lambda_i^2$ is applied when 
$\mathcal{C}$ is inverted.  With this feature in mind, we define the 
signal-to-noise ratio as
\begin{equation}\label{eq:s2n}
  \left(\!\frac{S}{N}\!\right)_i = \left| \frac{ \widehat{Q}_i }{ \lambda_i } \right| \; .
\end{equation}
We note that this $S/N$ estimate is a \emph{lower} bound on the true $S/N$ due to the 
SVD. To remove noise and avoid numerical instabilities, we trim eigenmodes corresponding 
to low singular values.  \citet{GS05} suggest keeping eigenmodes resolved better than 
the sampling error in the covariance matrix.  Since our covariance matrices are normalized 
(i.e. the diagonal elements are set to one), the singular values are directly related to 
sampling error, and we require the so-called ``dominant modes'' \citep{GS05} to satisfy:
\begin{equation}\label{eq:svcut}
  \lambda^2_i > \sqrt{ 2 / N } \;, 
\end{equation}
where $N$ refers to the number of samples used to estimate the covariance matrix. 

The advantage to using this eigenmode analysis for fitting is threefold. 
First, it correctly incorporates the correlation between measurement bins.
Second, by performing the fit in the rotated basis of the eigenmodes, the residuals of the 
fit are more Gaussian and the degrees of freedom are properly addressed 
(e.g. 3 eigenmodes really only fits over 3 numbers).  
Finally, using only dominant modes removes artifacts due to noise in the estimated 
covariance matrices. 
For example, when using the full covariance but not trimming any modes, noise can 
cause a fit to converge on incorrect values with artificially small errors 
(and falsely high $S/N$).  This effect becomes worse as the covariance becomes 
less resolved.  
Conversely, fitting over dominant eigenmodes helps to eliminate any problems from 
unresolved parts of the error estimation \citep[ Figure 13]{GS05}, and has the benefit 
of dealing with singular covariance matrices.

\section{Galaxy-Mass Bias}
\label{s:bias}

\begin{figure}
  \centering
  \includegraphics[angle=0,width=\linewidth]{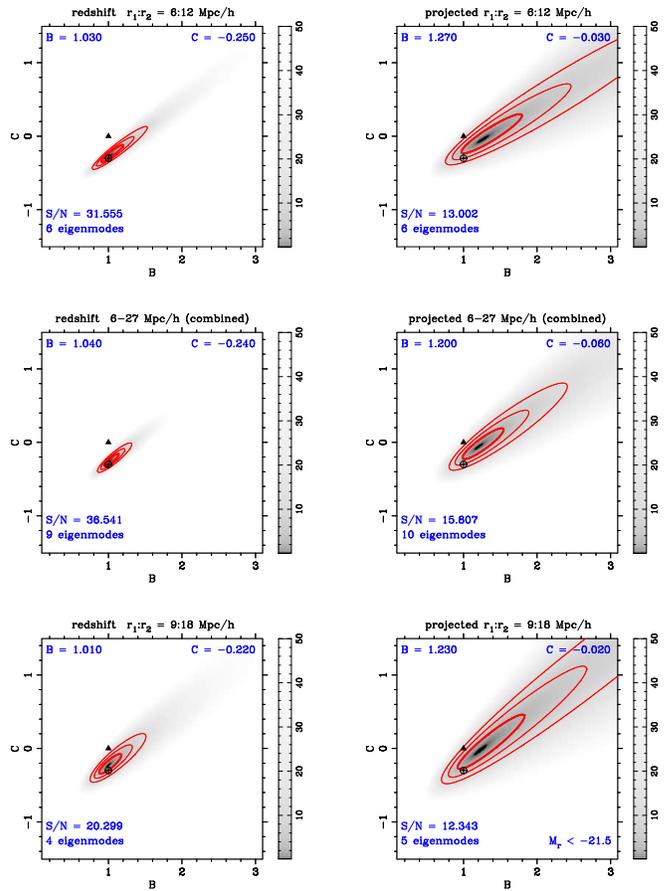}
  \caption[Galaxy Mass Bias Constraints: $ M_r < -21.5 $]{ 
    Constraints on the galaxy-mass bias parameters using the $M_r < -21.5$ galaxy sample and the HV 
    simulation for mass estimates.  
    The left column corresponds to fits using $Q_z(\theta)$ (redshift space) with the right column 
    fit using $Q_{proj}(\theta)$ (projected space).
    The top and bottom panels represent individual fits with triangles of $r_1 = 6 $ and $9 \; \hmpc$ as indicated.  
    The middle panels are a joint fit using both triangles.  
    There are two points of comparison marked: an unbiased result with $(B=1.0;C=0.0)$ and only non-linear bias $(B=1.0;C=-0.3)$.  
    The contours denote the $1, 2$ and $3 \sigma$ levels from the $\Delta \chi^2$ distribution of two parameters. 
  } 
  \label{f:bias_mt21_5} 
\end{figure}

\begin{figure}
  \centering
  \includegraphics[angle=0,width=\linewidth]{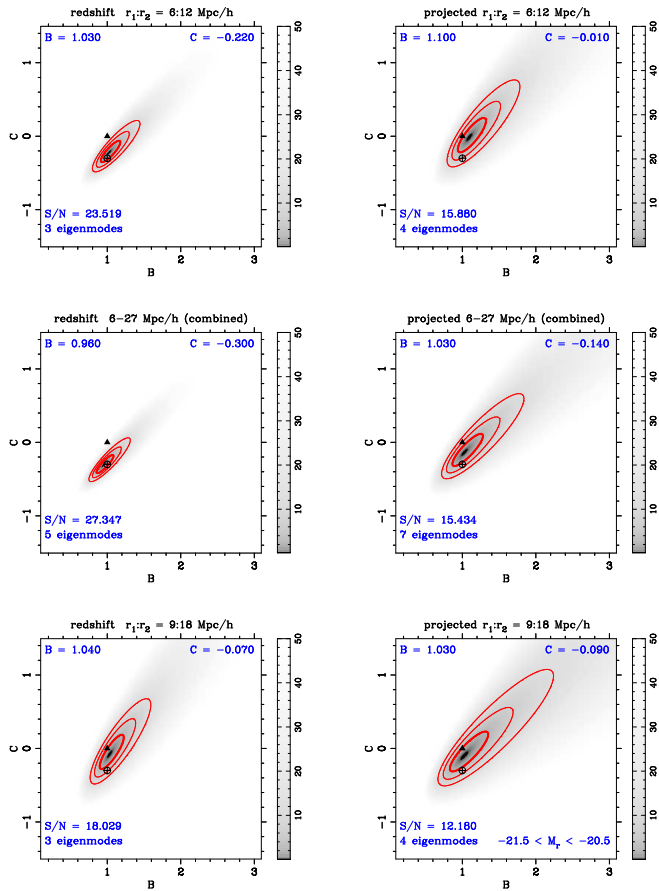} 
  \caption[Galaxy Mass Bias Constraints: $-21.5 < M_r < -20.5$]{
    Analogous to Figure~\ref{f:bias_mt21_5}, but for $-21.5 < M_r < -20.5$ galaxies. 
  } 
  \label{f:bias_mb20_5} 
\end{figure}

\begin{figure}
  \centering
    \includegraphics[angle=270,width=\linewidth]{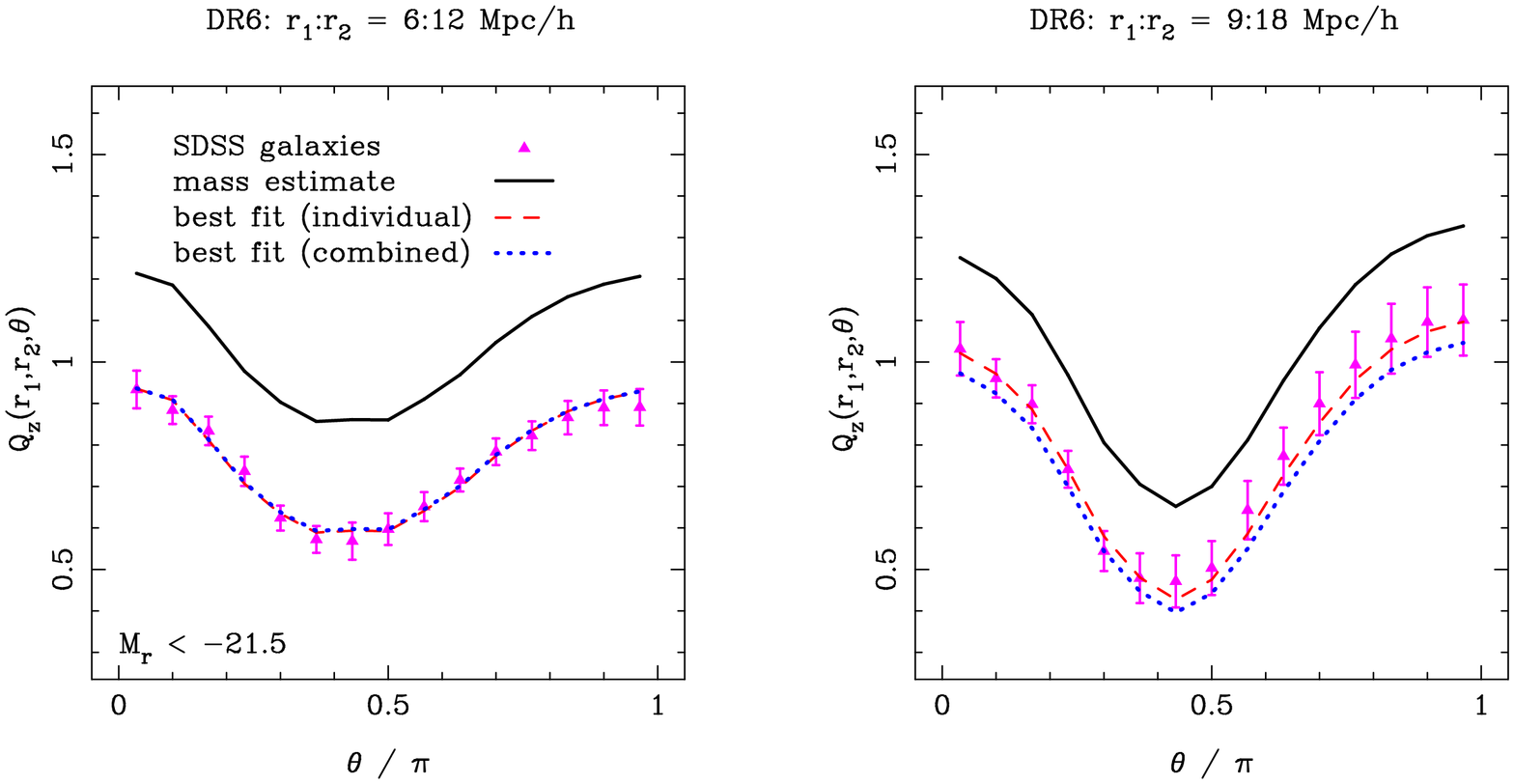}
    \includegraphics[angle=270,width=\linewidth]{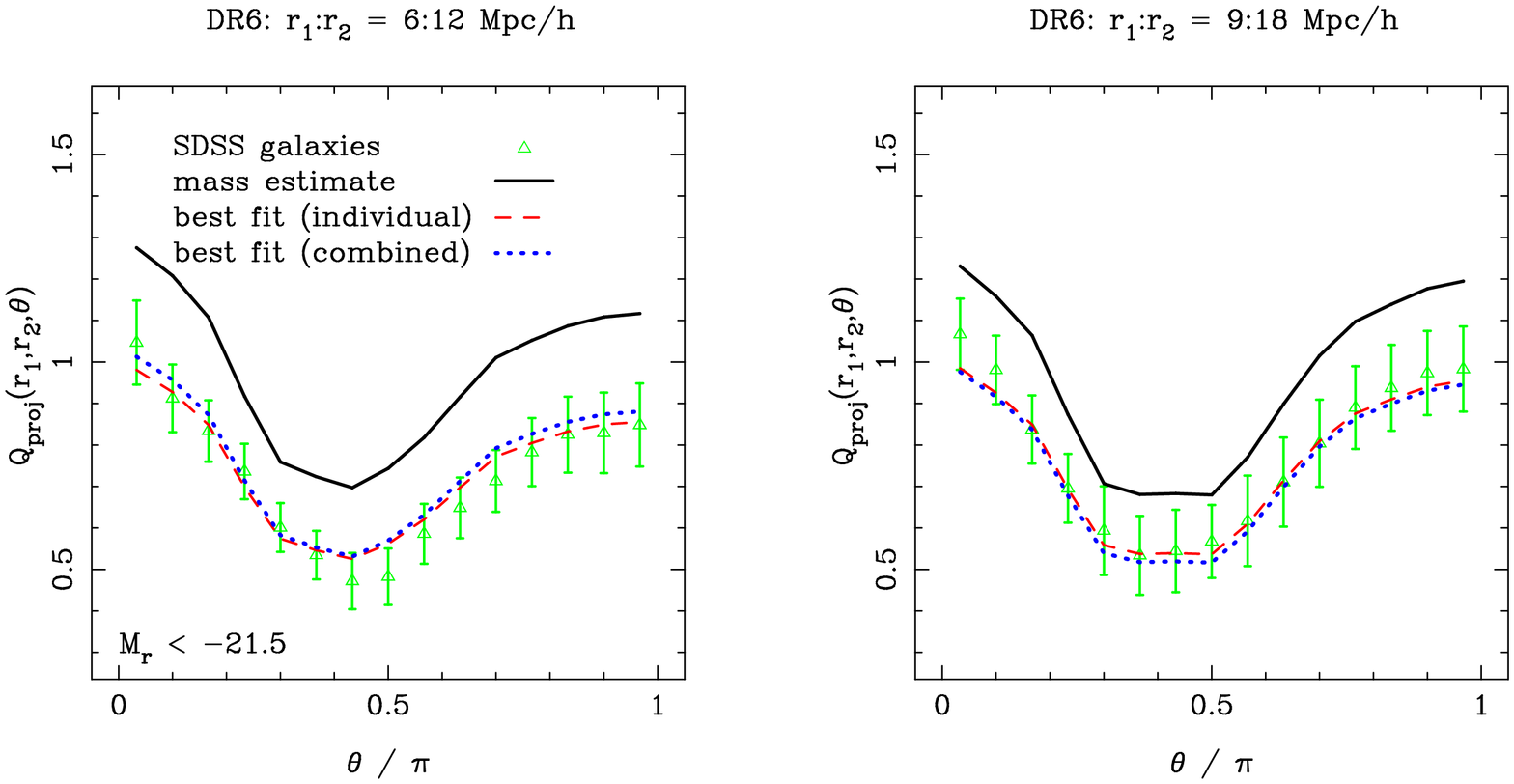}
  \caption[SDSS 3PCF with Best Fit Bias Parameters: $M_r < -21.5$]{ 
    The reduced 3PCF for the $M_r < -21.5$ sample showing the mass scaled to the ``best fit'' 
    galaxy-mass bias parameters.  The top two panels correspond to redshift space, and the bottom two 
    to projected space.  From left to right, the scale of the triangle increases as noted.  The red 
    (dashed) line represents an individual fit only to that triangle scale, and the blue (dotted) line 
    shows a joint fit between both scales. 
  } 
  \label{f:Qfit_mt21_5} 
\end{figure}

\begin{figure}
  \centering
    \includegraphics[angle=270,width=\linewidth]{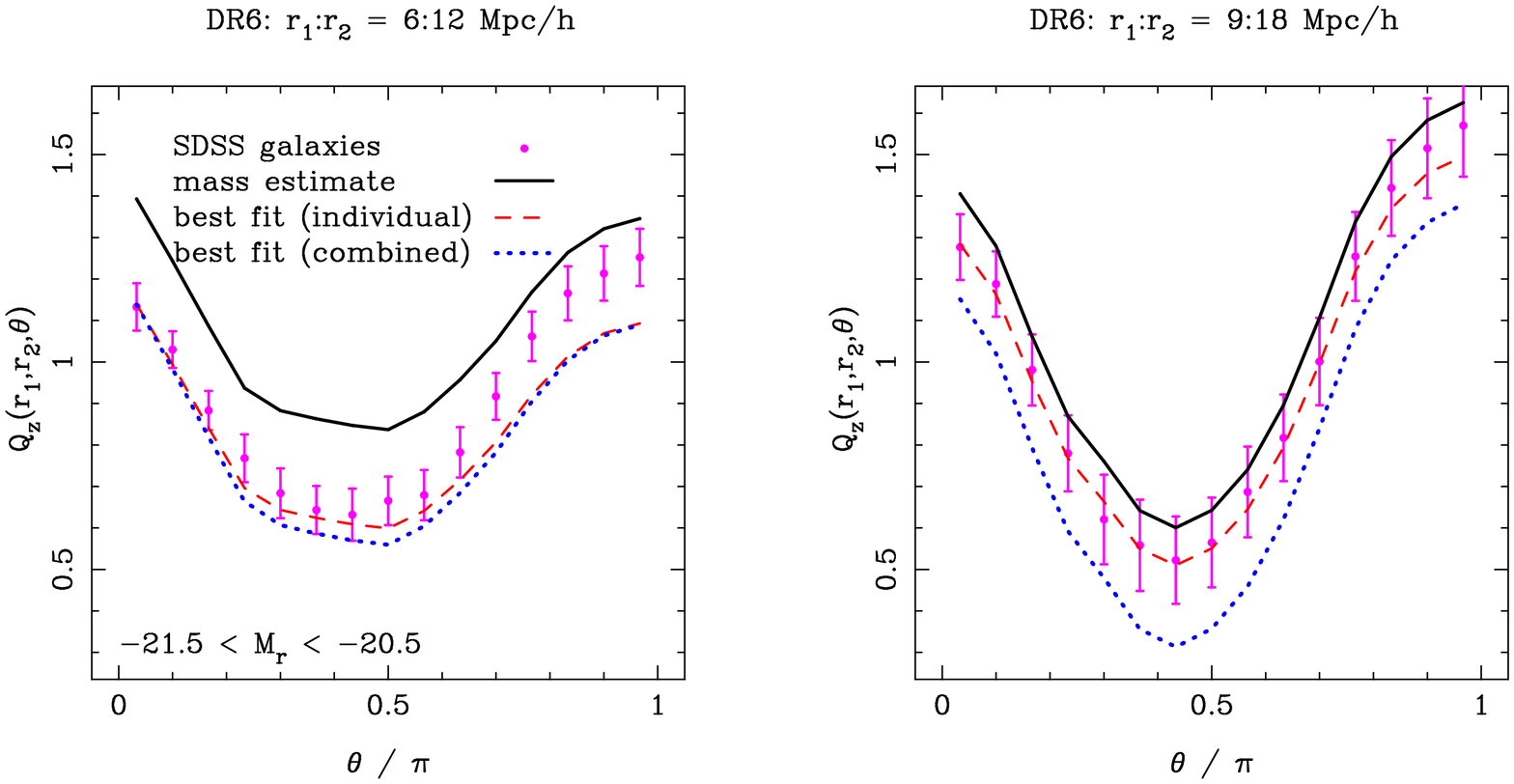}
    \includegraphics[angle=270,width=\linewidth]{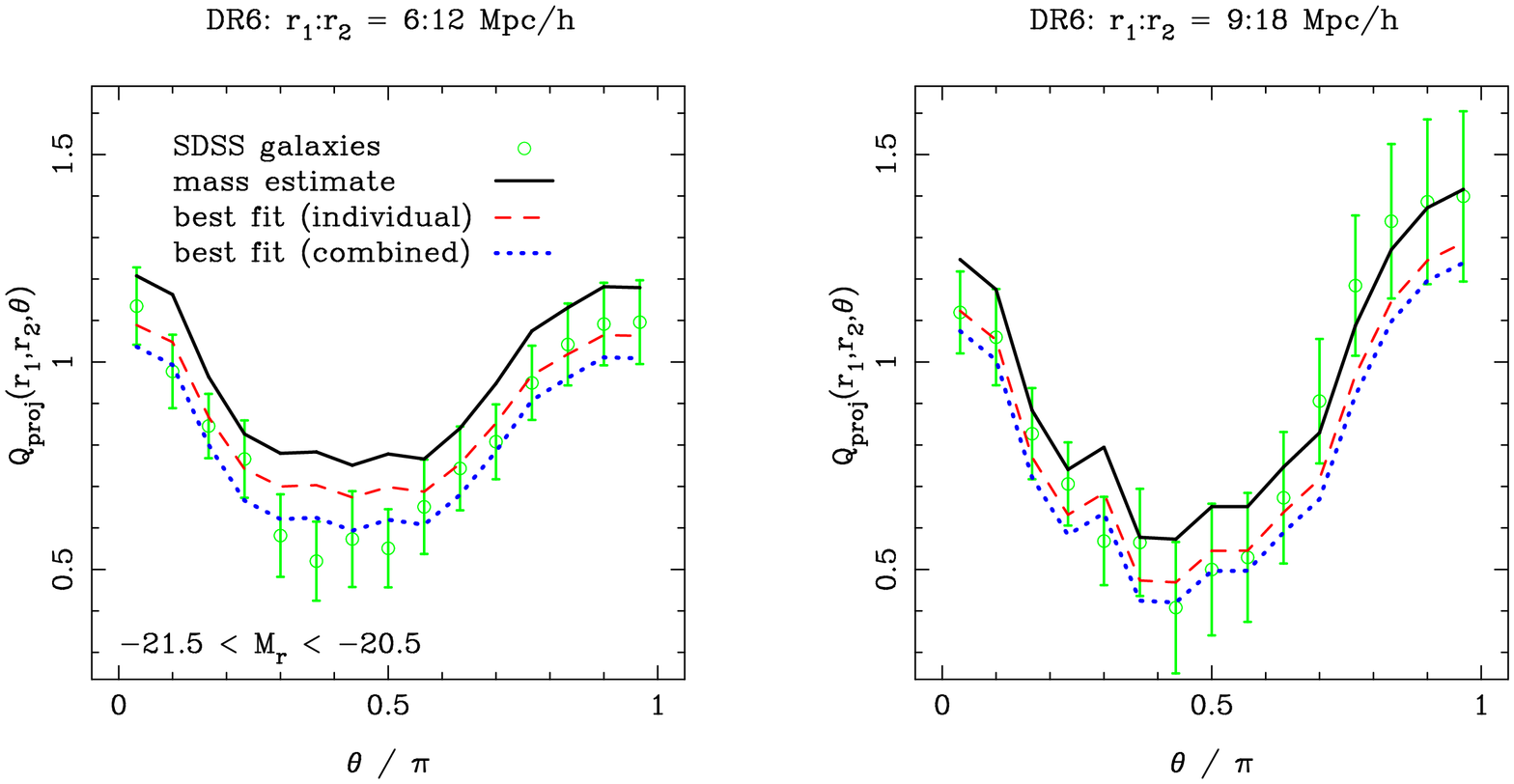}
  \caption[SDSS 3PCF with Best Fit Bias Parameters: $-21.5 < M_r < -20.5$]{
    Like Figure~\ref{f:Qfit_mt21_5} but for the $-21.5 < M_r < -20.5$ sample.  The reduced 3PCF showing 
    the mass scaled to the ``best fit'' galaxy-mass bias parameters.  The top two panels correspond to 
    redshift space, and the bottom two to projected space.  From left to right, the scale of the triangle 
    increases as noted.  The red (dashed) line represents an individual fit only to that triangle scale, 
    and the blue (dotted) line shows a joint fit between both scales.  
  } 
  \label{f:Qfit_mb20_5} 
\end{figure}

We want to constrain the galaxy-mass bias described by \eqref{eq:bias_mod} using the full 
configuration dependence of the reduced 3PCF in the quasi-linear regime.  For the galaxy data, 
we use measurements in both redshift space, $Q_z(\theta)$, and projected space, $Q_{proj}(\theta)$
as presented in \citet{mcbride:10}.  We estimate the bias parameters by comparing to mass 
estimates obtained from dark matter particles in the HV simulation (see \S{ss:hv}).
We expect redshift distortions to affect the bias relation, which we partially neglect 
\citep{scoccimarro:99}.  In particular, we account for the effects of redshift distortions 
by applying a distortion distance to the dark matter particles based on their velocities 
for our mass measurement.  However, this is not completely sufficient as redshift 
distortions alter the bias relation in \eqref{eq:biasQ}, especially for $Q_z$ 
\citep{scoccimarro:99,scoccimarro:01}.  
We expect that $Q_{proj}$ will be predominantly unaffected and roughly equivalent 
to real space measurements for this parameterization \citep[see e.g.][]{zheng:04}. 

We restrict our analysis to scales above $6 \; \hmpc$, corresponding 
to $Q(r_1, r_2, \theta)$ with $r_1 = 6$ and $9 \, \hmpc$ with configurations 
having $r_2 = 2 r_1$.
We let $\theta$ vary between $0$ and $\pi$ using $15$ bins, as detailed in 
\citet{mcbride:10}.
We investigate galaxy-mass bias in two samples where the covariance is well determined: 
BRIGHT ($M_r < -21.5$) and LSTAR ($-21.5 < M_r < -20.5$) as listed in 
Table~\ref{t:gal_samples}. 
We remove the least significant eigenmodes during the fit by applying the criteria in 
\eqref{eq:svcut}. We discuss possible effects of using a different number of modes in 
\S\ref{s:errors}. For each galaxy sample, we perform six independent fits: a series of 
three different scales for measurements in both redshift and projected space.  We use the full 
configuration dependence for triangles with $r_1 = 6 $ and $9 \; \hmpc$, as well as a joint fit 
using both scales. For the joint fit, we estimate the full combined covariance matrix to correctly 
account for overlap and correlation and use the same eigenmode analysis.  This changes the number 
of available modes from $15$ in the individual fits to $30$ modes for the combined joint fit.
We estimate the covariance for these samples using $30$ jackknife samples, where our 
our jackknife regions have equal unmasked area on the sky and use the full redshift 
distribution of the observational galaxy sample \citep{mcbride:10}.

\subsection{Constraining Non-Linear Local Bias}
\label{ss:nlbias}

We constrain the galaxy-mass bias using a maximum likelihood approach by calculating a 
simple $\chi^2$ statistic where the likelihood $\mathcal{L} \propto \exp( -\chi^2 / 2 ) $ and 
\begin{eqnarray} \label{eq:chi2}
  \chi^2      &=&       \vec{\Delta}^T \bm{\mathcal{C}}^{-1} \vec{\Delta} \; , \nonumber \\
  \Delta_i    &=&       \frac{Q_i - Q_i^{(t)}}{ \sigma_i } \; .
\end{eqnarray}
We determine the theoretical model, $Q^{(t)}$, by scaling the mass measurement from the 
HV simulation, $Q_m$, with bias parameters $B$ and $C$ as per \eqref{eq:biasQ}.  We evaluate 
$\mathcal{L}$ on a grid using the ranges: $B = 0.1 \ldots 3.0$ and $C = -1.5 \ldots 1.5$ 
with a step-size of $0.01$. We tested for discrepancies using a factor of $10$ finer 
spacing between grid elements with no significant differences to the fitted results.

We first examine the BRIGHT sample ($M_r < -21.5$), with the likelihood space of the six 
2-parameter fits displayed in Figure~\ref{f:bias_mt21_5}.  We include contours for Gaussian 
$1,2$ and $3\sigma$ levels which identify regions of probabilities for $68.3, 95.5$ and $99.7$\%. 
We calculate these from the $\Delta\chi^2$ distribution for a 2-parameter fit 
(i.e. two degrees-of-freedom), with corresponding values of 2.3, 6.2, and 11.8 
from the best fit value.
We include two reference points for comparison, the \emph{unbiased} 
result where $(B=1.0;C=0.0)$ along with a potential negative quadratic bias term accounting for 
the entire galaxy bias $(B=1.0;C=-0.3)$ \citep[similar comparison to Figure 5 in][]{gaztanaga:05}.

We can clearly see the degeneracy between $B$ and $C$ in Figure~\ref{f:bias_mt21_5}, visible as 
the elongated diagonal contour. Larger values of $B$ remain likely with larger values of $C$, 
consistent with our expectation of degeneracy by inspecting the bias relation in \eqref{eq:biasQ}. 
The size of the errors are notably larger for projected space measurements, as well as lower values 
for the overall $S/N$.  This results from the larger uncertainties in the projected measurements 
\citep{mcbride:10}.  Since the scale $r_p$ represents a projection that incorporates larger scales 
(determined by the line-of-sight integration $\pi_{max}$), projected measurements are more 
sensitive to the dominant uncertainty from cosmic variance that increases with scale.  In all cases, 
the \emph{unbiased} $(B=1;C=0)$ model is excluded at greater than a $2\sigma$ level.  To see 
the success of the fit ``by eye'', we plot the 3PCF for dark matter, galaxies and best fit scaled 
model for this sample in Figure~\ref{f:Qfit_mt21_5}.  Both the ``individual'' fits and ``combined'' 
joint fit produce models that well match the data.

Next, we fit the galaxy-mass bias parameters using the LSTAR sample ($-21.5 < M_r < 
-20.5$). This sample spans a unit bin in magnitude, and consists only of galaxies fainter 
than the previous bright sample.  The results of the fit with likelihood contours are 
shown in Figure~\ref{f:bias_mb20_5}.  The uncertainties appear reduced in size -- a 
striking difference with respect to the BRIGHT sample in Figure~\ref{f:bias_mt21_5}.  In 
addition, the slope of the ``line of degeneracy'' between $B$ and $C$ has shifted.  We 
reason that this is in part due to the increased statistical significance of the larger sample, 
as both the measurements and covariance are better resolved.  Due to the higher number 
density of galaxies, we re-measured the 3PCF using a finer binning scheme (fractional 
bin-width of $f=0.1$ as opposed to $f=0.25$, see comparison in Appendix~\ref{s:binning}).  
With the finer binning, we see a stronger configuration dependence, which will alter the 
degeneracy between $B$ and $C$.  We note that many of the best fit $B$ values appear 
smaller, which we expect for a fainter sample \citep{zehavi:05,zehavi:10}.  The same line 
of reasoning suggests that the ``unbiased'' model $(B=1;C=0)$ should be more likely to 
fit.  

As before, we plot the respective best fit model in comparison with the dark matter and galaxy 3PCF in 
Figure~\ref{f:Qfit_mb20_5}.  There is a smaller difference between HV (mass) and galaxy measurements, 
as this sample is fainter.  We notice some noise of the HV measurement for $Q_{proj}$, making the model 
not quite as smooth.  We note that by eye, $Q_z$ on larger scales indicates a slight bias for the 
combined fit, with the model undershooting the data and $1\sigma$ uncertainties.  Significant off-diagonal 
structure in the covariance matrix can produce a fit where ``chi-by-eye'' suggests a poor fit.  
Since the $r_1 = 6 \; \hmpc$ measurements in $Q_z$ have much smaller errors, these scales drive 
the fit making measurements with $r_1 = 9 \; \hmpc$ appear a poor match to the ``best fit'' 
model.

We summarize the results of our two parameter constraints for the BRIGHT and LSTAR sample 
in Table~\ref{t:bias}.  The BRIGHT sample ($M_r < -21.5$) represents galaxies with $r$-band magnitudes 
significantly brighter than $\Lstar$ where $M_r \sim -20.4$ \citep{blanton:03}.  We typically consider 
$\Lstar$ galaxies to have a linear bias, i.e. where $B \sim 1$, and we might expect this brighter 
sample to have a larger $B$ value.  The constraints from projected measurements appear to follow this 
logic;  the best fit values on $Q_{proj}$ in the fainter LSTAR sample ($-21.5 < M_r < -20.5$) 
are lower with $B \sim 1$.  Redshift space measurements, $Q_z$, appear consistent with 
$B \sim 1$ for all fits, but at the same time values of $C$ are lower, reflecting the degeneracy of 
the $B$ and $C$ parameters.  The reduced $\chi^2_{\nu}$ values show an acceptable fit in almost all cases;
the exceptions are the two $Q_z$ fits using the $r_1 = 6 \; \hmpc$ triangles for the LSTAR sample.  
Consequently, the joint fit appears to be the poorest match in Figure~\ref{f:Qfit_mb20_5}.  
The \Dchisq\ in Table~\ref{t:bias} displays the likelihood an unbiased model is from the best 
fit parameters.  We find an unbiased model is ruled out for the BRIGHT galaxy sample at greater 
than $4.8\sigma$ in redshift space and $2.6\sigma$ in projected space.  We cannot conclude the same for the 
LSTAR sample, which is largely consistent with an unbiased model.  We generally consider bright 
galaxies to be more biased \citep{zehavi:05,zehavi:10}. The LSTAR sample is a magnitude 
bin around $\Lstar$ and fainter than the BRIGHT sample, and we expect a better consistency with the 
unbiased model.

\begin{table*}
  \centering
  \begin{tabular}{lcccccc}
    \hline
    \hline
    \multicolumn{7}{c}{\bfseries Galaxy-Mass Bias Parameters from SDSS}  \\
    \hline
     Measurement & Scales ($\hmpc$) & B & C & $\chi^2_\nu$ & D.o.F. & unbiased $\Delta\chi^2$ \\ 
    \hline
    \hline
     BRIGHT-z    & 6-18 & $ 1.03_{-0.08}^{+0.11} $ & $ -0.25_{-0.06}^{+0.08} $ & 1.48 & 6-2 &  $118.86$ ($10.7\sigma$) \\
     BRIGHT-proj & 6-18 & $ 1.27_{-0.21}^{+0.30} $ & $ -0.03_{-0.14}^{+0.19} $ & 0.78 & 6-2 &  $ 16.43$ ($ 3.6\sigma$) \\
     BRIGHT-z    & 6-27 & $ 1.04_{-0.06}^{+0.06} $ & $ -0.24_{-0.05}^{+0.05} $ & 0.83 & 9-2 &  $132.54$ ($11.3\sigma$) \\
     BRIGHT-proj & 6-27 & $ 1.20_{-0.14}^{+0.21} $ & $ -0.06_{-0.11}^{+0.15} $ & 0.45 & 10-2 & $ 18.70$ ($ 3.9\sigma$) \\
     BRIGHT-z    & 9-27 & $ 1.01_{-0.09}^{+0.10} $ & $ -0.22_{-0.08}^{+0.09} $ & 0.60 & 4-2 &  $ 26.90$ ($ 4.8\sigma$) \\
     BRIGHT-proj & 9-27 & $ 1.23_{-0.22}^{+0.34} $ & $ -0.02_{-0.18}^{+0.27} $ & 0.34 & 5-2 &  $  9.44$ ($ 2.6\sigma$) \\
    \hline
     LSTAR-z     & 6-18 & $ 1.03_{-0.07}^{+0.09} $ & $ -0.22_{-0.08}^{+0.10} $ &13.47 & 3-2 &  $ 28.07$ ($ 4.9\sigma$) \\
     LSTAR-proj  & 6-18 & $ 1.10_{-0.11}^{+0.13} $ & $ -0.01_{-0.14}^{+0.16} $ & 0.85 & 4-2 &  $  3.00$ ($ 1.2\sigma$) \\
     LSTAR-z     & 6-27 & $ 0.96_{-0.07}^{+0.08} $ & $ -0.30_{-0.08}^{+0.08} $ & 3.22 & 5-2 &  $ 45.85$ ($ 6.5\sigma$) \\
     LSTAR-proj  & 6-27 & $ 1.03_{-0.11}^{+0.15} $ & $ -0.14_{-0.12}^{+0.16} $ & 1.07 & 7-2 &  $  5.86$ ($ 1.9\sigma$) \\
     LSTAR-z     & 9-27 & $ 1.04_{-0.09}^{+0.11} $ & $ -0.07_{-0.14}^{+0.16} $ & 0.07 & 3-2 &  $  2.37$ ($ 1.0\sigma$) \\
     LSTAR-proj  & 9-27 & $ 1.03_{-0.13}^{+0.19} $ & $ -0.09_{-0.15}^{+0.19} $ & 1.75 & 4-2 &  $  1.93$ ($ 0.9\sigma$) \\
    \hline
  \end{tabular}
  \caption[Galaxy-Mass Bias Parameters from SDSS DR6]{ 
    The two-parameter best fit galaxy-mass bias parameters, using \eqref{eq:biasQ} with the 
    configuration dependence in the reduced 3PCF from SDSS DR6 galaxy samples in comparison with 
    dark matter clustering from the Hubble volume simulation.  
    The fits are performed separately on two galaxy samples BRIGHT ($M_r < -21.5$) and LSTAR 
    ($-21.5 < M_r < -20.5$) using measurements in redshift space (denoted with ``z'') as well as 
    projected space (``proj'').  
    The second column lists the range of scales used for the respective fit.  
    The errors are marginalized $1\sigma$ bounds calculated by the range within $\Delta\chi^2 \le 1$ from the best 
    fit value.
    The quality of the best fit value is stated with the reduced chi-square $\chi^2_{\nu} = \chi^2 / \text{D.o.F}$.  
    The degrees of freedom (D.o.F.) correspond to the number of eigenmodes used minus the number of 
    parameters ($2 for all these fits$).
    The last column lists the $\Delta\chi^2$ value to quantify the likelihood of an ``unbiased'' model 
    matching the data, i.e. $(B=1;C=0)$, with a likelihood expressed in the number of $\sigma$ from 
    by the standard Gaussian assumption for the $\Delta\chi^2$ distribution.
  }
  \label{t:bias}
\end{table*}

\begin{table*}
  \centering
  \begin{tabular}{lccccc}
    \hline
    \hline
    \multicolumn{6}{c}{\bfseries Galaxy-Mass Bias without Quadratic Term }  \\
    \hline
     Measurement & Scales ($\hmpc$) & B & $\chi^2_\nu$ & D.o.F. & $\Delta\chi^2$ from best fit \\ 
    \hline
    \hline
     BRIGHT-z    & 6-18 & $ 1.34_{-0.04}^{+0.04} $ & 2.42 & 6-1 &  $  6.16$ ($ 2.0\sigma$)\\ 
     BRIGHT-proj & 6-18 & $ 1.31_{-0.09}^{+0.11} $ & 0.63 & 6-1 &  $  0.04$ ($ 0.0\sigma$)\\ 
     BRIGHT-z    & 6-27 & $ 1.30_{-0.03}^{+0.03} $ & 2.23 & 9-1 &  $ 12.01$ ($ 3.0\sigma$)\\ 
     BRIGHT-proj & 6-27 & $ 1.27_{-0.07}^{+0.09} $ & 0.42 & 10-1 & $  0.16$ ($ 0.1\sigma$)\\ 
     BRIGHT-z    & 9-27 & $ 1.24_{-0.06}^{+0.06} $ & 1.85 & 4-1 &  $  4.35$ ($ 1.6\sigma$)\\ 
     BRIGHT-proj & 9-27 & $ 1.25_{-0.09}^{+0.11} $ & 0.26 & 5-1 &  $  0.01$ ($ 0.0\sigma$)\\ 
    \hline                                                                                 
     LSTAR-z     & 6-18 & $ 1.21_{-0.04}^{+0.05} $ & 8.68 & 3-1 &  $  3.90$ ($ 1.5\sigma$)\\ 
     LSTAR-proj  & 6-18 & $ 1.11_{-0.06}^{+0.07} $ & 0.57 & 4-1 &  $  0.01$ ($ 0.0\sigma$)\\ 
     LSTAR-z     & 6-27 & $ 1.23_{-0.04}^{+0.04} $ & 4.59 & 5-1 &  $  8.70$ ($ 2.5\sigma$)\\ 
     LSTAR-proj  & 6-27 & $ 1.15_{-0.07}^{+0.07} $ & 1.02 & 7-1 &  $  0.76$ ($ 0.4\sigma$)\\ 
     LSTAR-z     & 9-27 & $ 1.08_{-0.05}^{+0.06} $ & 0.14 & 3-1 &  $  0.21$ ($ 0.1\sigma$)\\ 
     LSTAR-proj  & 9-27 & $ 1.11_{-0.08}^{+0.09} $ & 1.25 & 4-1 &  $  0.25$ ($ 0.1\sigma$)\\ 
    \hline
  \end{tabular}
  \caption[Galaxy-Mass Bias without a Quadratic Term]{ 
    The single-parameter best fits for galaxy-mass bias using \eqref{eq:biasQ} where we constrain 
    $C=0$.  We fit the configuration dependence of the reduced 3PCF from SDSS DR6 galaxy samples in 
    comparison with dark matter clustering from the Hubble volume simulation.  
    Fits are performed separately on two galaxy samples BRIGHT ($M_r < -21.5$) and LSTAR 
    ($-21.5 < M_r < -20.5$) using measurements in redshift space (denoted with ``z'') as well as 
    projected space (``proj'').  
    The second column lists the range of scales used for the respective fit.  
    The errors are marginalized $1\sigma$ bounds calculated by the range within $\Delta\chi^2 \le 1$ from the best 
    fit value.
    The quality of the best fit value is stated with the reduced chi-square $\chi^2_{\nu} = \chi^2 / \text{D.o.F}$.  
    The degrees of freedom (D.o.F.) correspond to the number of eigenmodes used minus the number of 
    parameters (in this case, just one).
    The last column lists the $\Delta\chi^2$ value to quantify the difference in likelihood of this 
    model with $C=0$ compared with the best fit of a two-parameter fit (i.e. Table~\ref{t:bias}).
  }
  \label{t:linbias}
\end{table*}

\subsection{Non-zero Quadratic Bias?}
\label{ss:linbias}

With our two parameter likelihood space, we can investigate the statistical significance of 
a non-zero quadratic bias term ($b_2$ which is encapsulated in $C = b_2 / b_1$).  We use 
the same configuration dependence of the 3PCF, and the measured covariance, but restrict the two 
parameter fit such that $C = 0$.  We evaluate the best fit $B$, the quality of the fit 
(via the reduced $\chi^2_\nu$) as well as the $\Delta\chi^2$ for the best two-parameter 
fit, which we present in Table~\ref{t:linbias}.  For the BRIGHT sample, we notice the $B$ 
values are equivalent across both $Q_{proj}$ and $Q_{z}$ for the same scales on the same 
sample.  Since we removed the degeneracy (as $C$ is zero), this behavior makes sense and 
in agreement with the measurements.  We note that the typical $B$ values are larger for 
the BRIGHT sample, and lower for the fainter LSTAR sample. For $Q_{proj}$ our constraints 
find little statistical significance for a non-zero quadratic bias term; the likelihood 
difference is small and a linear bias term is sufficient to quantify the bias for both 
the BRIGHT and LSTAR samples.  Overall, measurements in redshift space ($Q_z$) more strongly 
suggest that $C \ne 0$, especially when using the smaller scale triangles ($r_1 = 6 \; \hmpc$).

\subsection{Relative Bias}
\label{ss:brel}

\begin{figure}
  \centering
  \includegraphics[angle=0,width=\plotwidth]{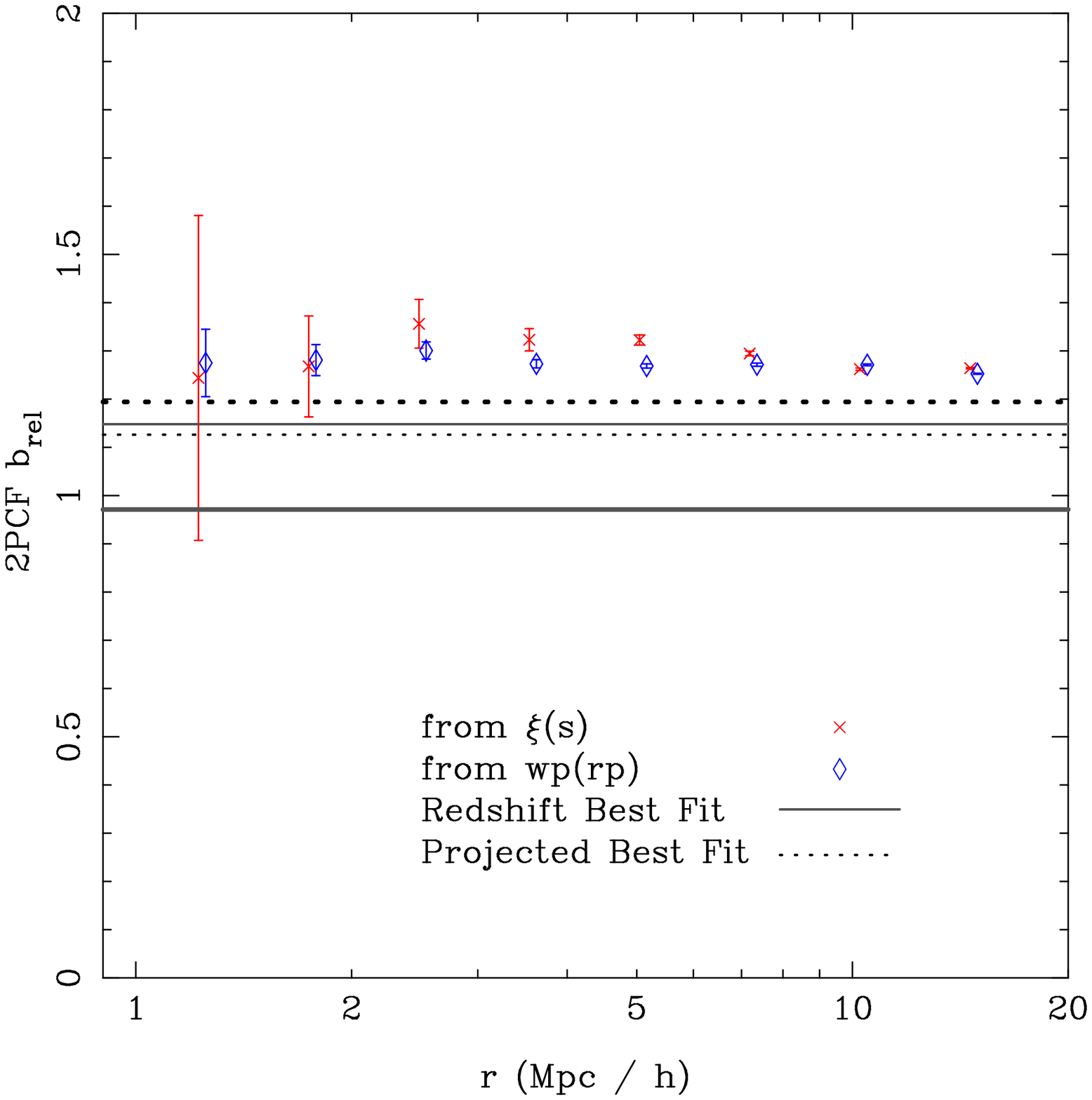}
  \caption[Relative Bias in 2PCF]{ 
    The relative bias $b^{(2)}_{rel} = \sqrt{\xi_{BRIGHT} / \xi_{LSTAR}}$ using measurements in 
    redshift (red 'x' symbols) and projected space (blue diamonds).  
    We calculate the uncertainties by propagating 1$\sigma$ values from the 2PCF.  
    The dotted and dashed lines display results from the best fit bias terms 
    at the largest scales ($9-27 \hmpc$).  
    The bold lines indicate values from the two-parameter fit (Table~\ref{t:bias}), and 
    the faint lines show the best linear fit (Table~\ref{t:linbias}).
  } 
  \label{f:brel_2pcf} 
\end{figure}

\begin{figure}
  \centering
  \includegraphics[angle=0,width=\plotwidth]{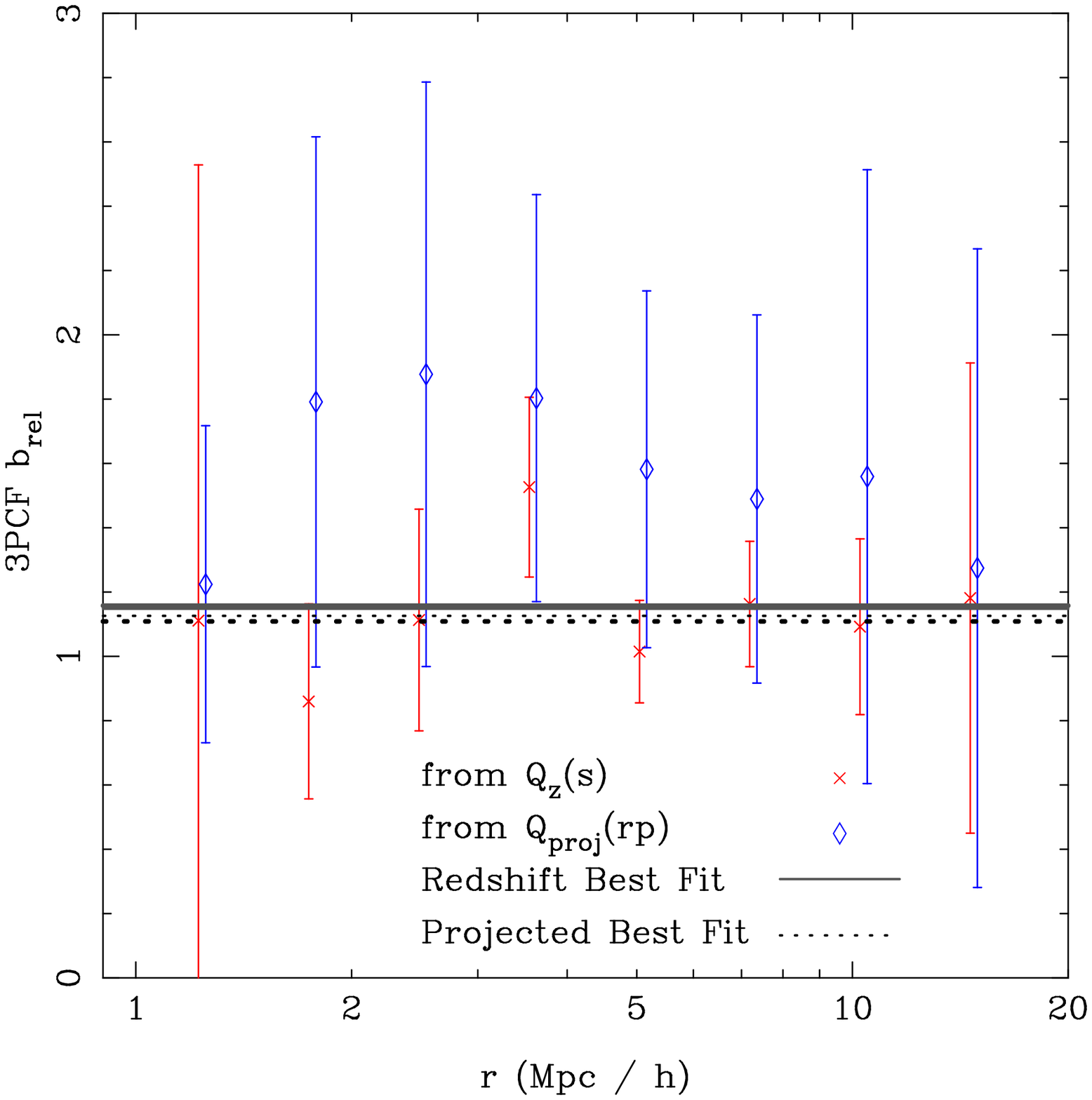}
  \caption[Relative Bias in $Q_{eq}(r)$]{ 
    Analogous to Figure~\ref{f:brel_2pcf} but for the 3PCF.
    The relative bias $b^{(3)}_{rel} = Q_{LSTAR} / Q_{BRIGHT}$ using measurements of equilateral 3PCF 
    in redshift (red 'x' symbols) and projected space (blue diamonds).  
    We calculate the uncertainties by propagating 1$\sigma$ values from the 3PCF.  
    The dotted and dashed lines display results using the best fit bias terms
    at the largest scales ($9-27 \hmpc$).
    The bold lines indicate values from the two-parameter fit (Table~\ref{t:bias}), and 
    the faint lines show the best linear fit (quadratic bias constrained to be zero; Table~\ref{t:linbias}).
  } 
  \label{f:brel_Qeq} 
\end{figure}

The \emph{relative} bias characterizes the relative clustering strength between different galaxy 
samples -- an alternative to the ``absolute'' galaxy-mass bias constrained previously.  Relative
bias is insensitive to cosmology and does not require assumptions to determine mass clustering.  
We can use the relative bias to check consistency with linear and quadratic bias parameters
obtained above in \S\ref{s:bias}.  For the 2PCF, the relative bias is simply:
\begin{equation} \label{eq:brel_2pcf}
  b^{(2)}_{rel} = \sqrt{ \frac{\xi_{BRIGHT}}{ \xi_{LSTAR} } } \; , 
\end{equation}
where $\xi$ can refer to redshift or projected space measurements.

We show $b^{(2)}_{rel}$ from the 2PCF in Figure~\ref{f:brel_2pcf}, using the linear bias parameters 
obtained from the best two-parameter fit (i.e. Table~\ref{t:bias}).  Both redshift space
and projected measurements agree and produce a flat relative bias, even at non-linear scales 
below a few $\hmpc$.  Two obvious discrepancies arise when comparing observational data to 
``best fit'' values.  First, neither redshift nor projected space fits appear to match data.  
Earlier we noted a substantial degeneracy between the linear and quadratic bias terms.  The
quadratic bias term is accounting for more of the clustering bias when we constrain with 
$Q(\theta)$, which isn't noticeable in the 2PCF.  This suggests we underpredict values of linear 
bias, either just for the BRIGHT sample or in unequal portions for both.  Second, there is a 
significant difference between these two estimates given the same galaxy samples, although the 
projected measurement appears closer to agreement.  

Let us consider the relative bias of the reduced 3PCF.  Since $Q$ is proportional to $1/B$, 
we define the relation 
\begin{equation} \label{eq:brel_Q}
  b^{(3)}_{rel} = \frac{ Q_{LSTAR} }{ Q_{BRIGHT} } \; .
\end{equation}
Figure~\ref{f:brel_Qeq} presents the relative bias of our DR6 galaxies for the equilateral 3PCF 
($Q_{eq}$) \citep{mcbride:10}. We note that $Q_{eq}$ is related but not identical to the 
configuration dependence measurements of $Q(\theta)$ used for constraining $B$ and $C$.  
Looking at Figure~\ref{f:brel_Qeq}, we see an obvious difference with respect to the 2PCF: the much 
larger uncertainties.  However, the predicted $b^{(3)}_{rel}$ from the galaxy-mass bias constraints appear 
much more consistent with the measurements, as opposed to the 2PCF.  The 3PCF results agree with the 
observational data, and show a much smaller discrepancy between redshift and projected space.  
The quadratic bias term ($C$) can properly account for the clustering difference that was missing 
in the relative bias of the 2PCF.  

\subsection{Implications for Cosmology: $\sigma_8$}
\label{ss:s8}

\begin{table}[b]
  \centering
  \begin{tabular}{lccc}
    \hline
    \hline
    \multicolumn{4}{c}{\bfseries Implied values of $\sigma_8$}  \\
    \hline
     Measurement & Scales ($\hmpc$) & B            & $\sigma_8$ \\
    \hline
    \hline
     BRIGHT-z    & 9-27 & $ 1.24_{-0.06}^{+0.06} $ & 0.96-1.13 \\
     BRIGHT-proj & 9-27 & $ 1.25_{-0.09}^{+0.11} $ & 1.02-1.12 \\ 
    \hline
     LSTAR-z     & 9-27 & $ 1.08_{-0.05}^{+0.06} $ & 0.88-0.97 \\ 
     LSTAR-proj  & 9-27 & $ 1.11_{-0.08}^{+0.09} $ & 0.83-0.97 \\ 
    \hline
  \end{tabular}
  \caption[Implied values of $\sigma_8$ from galaxy-mass bias]{ 
    We use galaxy-mass bias constraints from the configuration dependence of the 3PCF, $Q(\theta)$, 
    with measurements of the 2PCF, to estimate the implied values of $\sigma_8$ via \eqref{eq:bias2pt_s8}.  
    We use the largest triangle configurations for our two samples, and the 1-parameters constraints 
    on $B$.  The range of $\sigma_8$ does not represent formal uncertainties; we calculate 
    values from the range of uncertainties stated in $B$, neglecting additional errors from the 
    2PCF.  For reference, WMAP-5 (with SN and BAO) suggest $\sigma_8 = 0.82$ \citep{komatsu:09}.
  }
  \label{t:sigma8}
\end{table}

Better understanding galaxy-mass bias, or at the least accurately parameterizing it, allows one to
``calibrate out'' the effects of galaxies and infer properties of the underlying mass distribution
to constrain cosmology.  We can use our estimates of bias to probe the mass variance in spheres 
of $8 \; \hmpc$ radius, a common normalization of the amplitude of the matter power 
spectrum, $P(k)$.  The theoretical $\sigma_8$ is linearly extrapolated from a very early 
epoch until today, 
\begin{equation}\label{eq:sigma8}
  \sigma_8^2 = 4 \pi \int_0^\infty W^2(k,R = 8 \; \hmpc) P_{lin}(k) \frac{k^2 dk}{(2\pi)^3} \; ,
\end{equation}
where $W(k,R)$ is a top-hat window function in Fourier space for mode $k$ and smoothing radius $R$
and $P_{lin}(k)$ is the linear power spectrum.

In terms of our fitting formula on the 3PCF in \eqref{eq:biasQ}, we expand the bias relation for 
the 2PCF to highlight its dependence on $\sigma_8$
\begin{equation} \label{eq:bias2pt_s8}
  \xi_g(r) = B^2 \left(\frac{\sigma_8}{0.9}\right)^2 \xi_{m}(r) \; .
\end{equation}
Formally, the mass 2PCF already encodes a value of $\sigma_8$. As $\xi_m$ scales linearly with 
a change in the square of $\sigma_8$, we include an explicit scaling factor to account for a difference 
in $\sigma_8$ between the underlying mass of the observed galaxy distribution and that assumed in 
our estimate of mass clustering from \nbody\ results.  In our case, we use dark matter from the 
HV simulation where $\sigma_8 = 0.9$, explaining the denominator on the right hand side of 
\eqref{eq:bias2pt_s8}.  We can see that an incorrect assumption of $\sigma_8$ in the estimate of
mass will directly translate into a different value of the best $B$ describing galaxies.  Even if 
we use the above relation, \eqref{eq:bias2pt_s8}, $B$ and $\sigma_8$ are completely degenerate when 
solely considering the 2PCF.

By using the additional information available in the configuration dependence of the 
reduced 3PCF, we obtain a value of $B$ that is independent of $\sigma_8$, and breaks the degeneracy 
between the two parameters.  Formally, this is only true to leading order, as loop corrections 
in $Q(\theta)$ will add cosmological dependence which we neglect in this analysis.  

With an independent value of $B$ from \eqref{eq:biasQ}, we estimate $\sigma_8$ by utilizing the 
2PCF in \eqref{eq:bias2pt_s8}.  Ideally, we could construct a three-parameter fit to jointly 
constrain $B$, $C$ and $\sigma_8$ \citep[e.g.][ on 2dFGRS data]{pan:05}.  Or as an further extension, 
we could jointly fit over several samples, since they each have the same underlying $\sigma_8$.  
However, this additional complexity is beyond the scope of this analysis as our current uncertainties 
would yield poor constraints on $\sigma_8$.  We simply estimate the value of $\sigma_8$ implied by 
best fit bias parameters.  We restrict this estimate to the largest scale triangles 
($r_1 = 9 \; \hmpc$) to ensure we approach the linear regime (i.e. the scales we are most confident 
with using the local bias model).  Given our analysis of the relative bias, we use the 
larger $B$ values where we constrain $C=0$. We present these estimates in Table~\ref{t:sigma8}.

\section{Eigenvectors of the 3PCF Covariance Matrix}
\label{s:ev}

\begin{figure}
  \centering
  \includegraphics[angle=270,width=\hplotwidth]{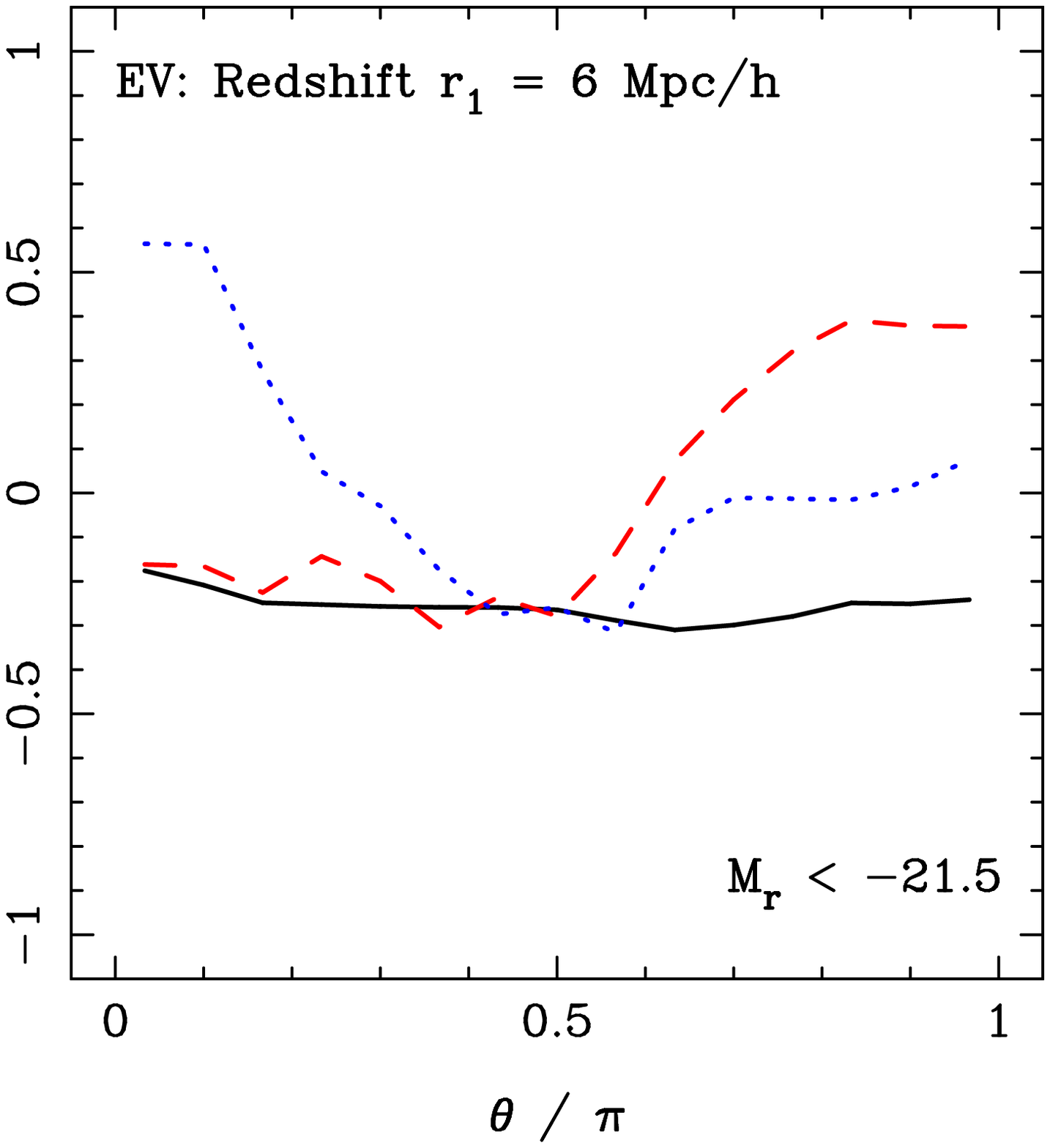}
  \includegraphics[angle=270,width=\hplotwidth]{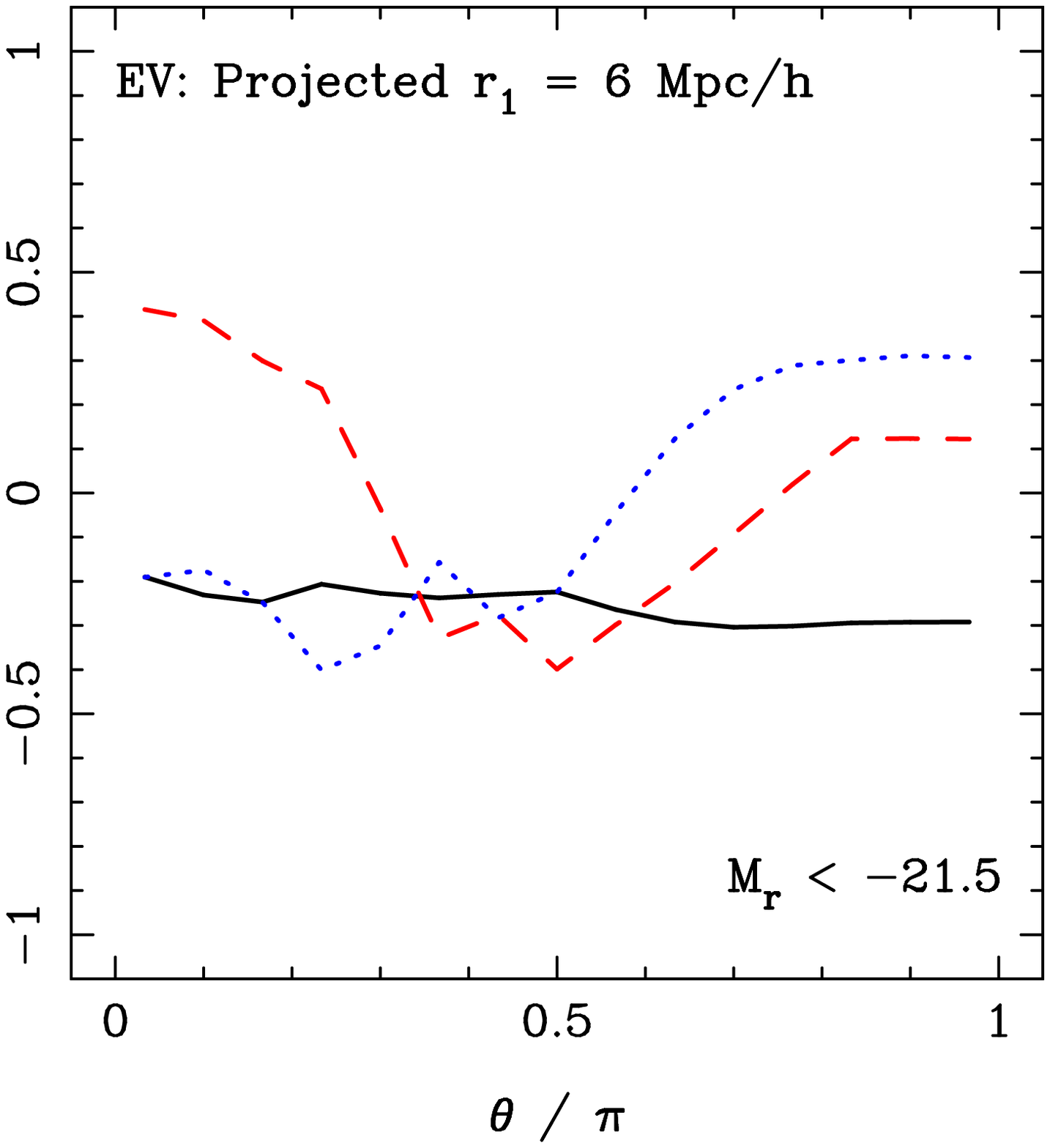}
  \includegraphics[angle=270,width=\hplotwidth]{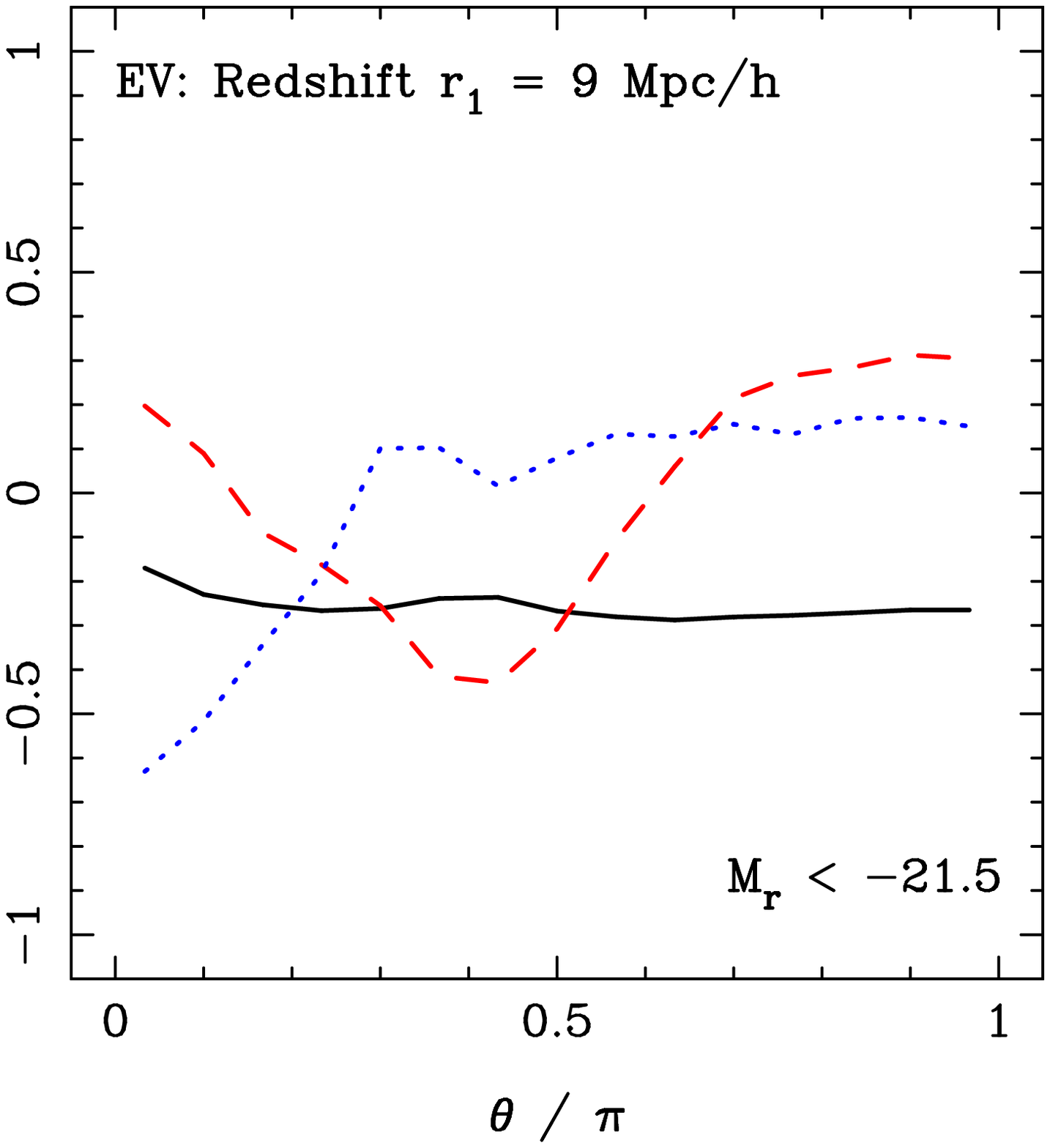}
  \includegraphics[angle=270,width=\hplotwidth]{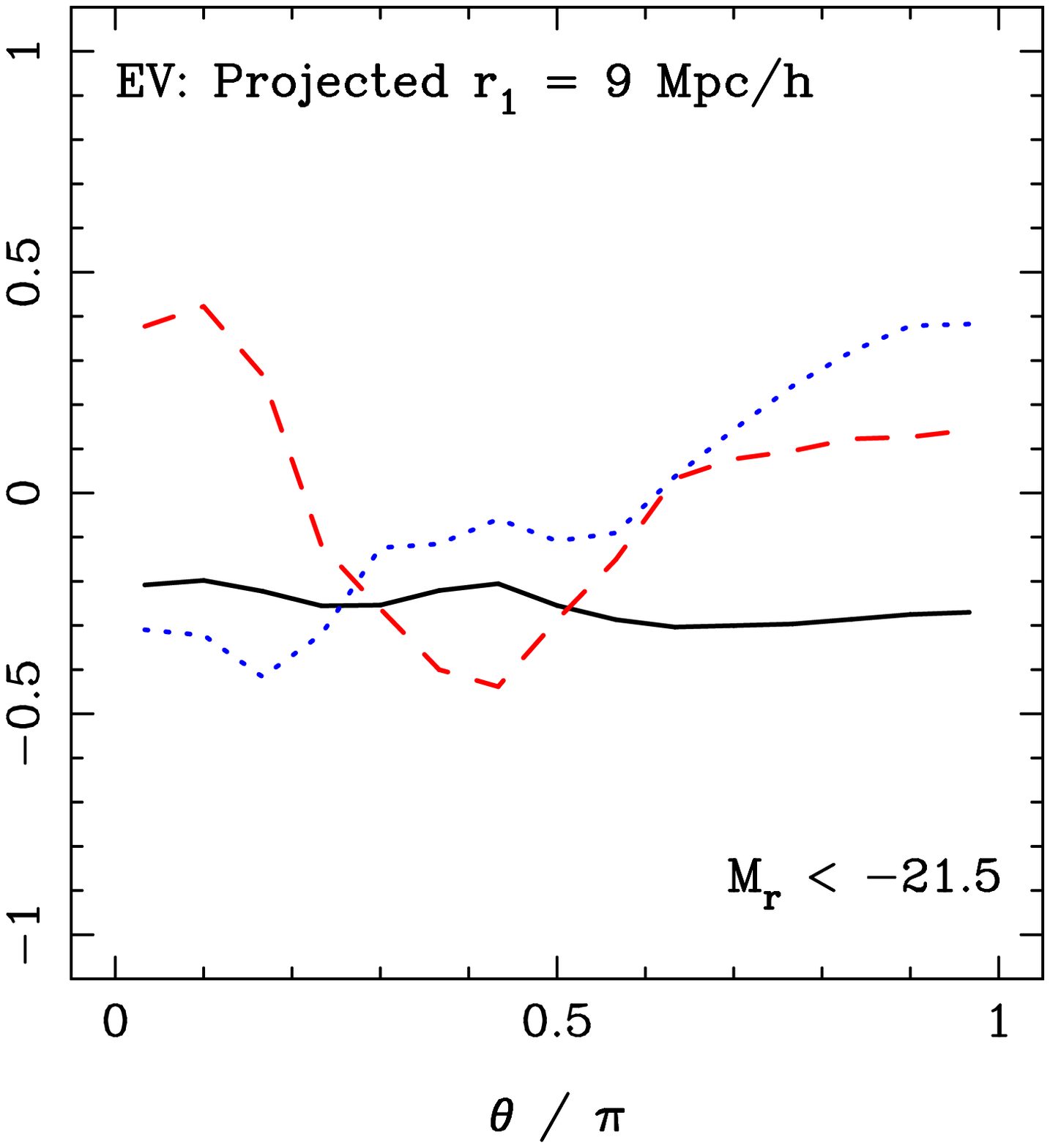}
  \caption[Top 3 Eigenvectors for $M_r < -21.5$]{ 
    Top three eigenvectors (EVs) chosen from the normalized covariance matrix in the 
    $M_r < -21.5$ galaxy sample. The sign of the EV is arbitrary.  The first EV (solid black) 
    shows equal weights for all bins.  The second (dashed red) and third (dotted blue) EV display 
    the configuration difference between perpendicular and co-linear triangles as well as the 
    scale variation as the scale of the third side increases.  
  } 
  \label{f:ev3_mt21_5} 
\end{figure}

\begin{figure}
  \centering
  \includegraphics[angle=270,width=\hplotwidth]{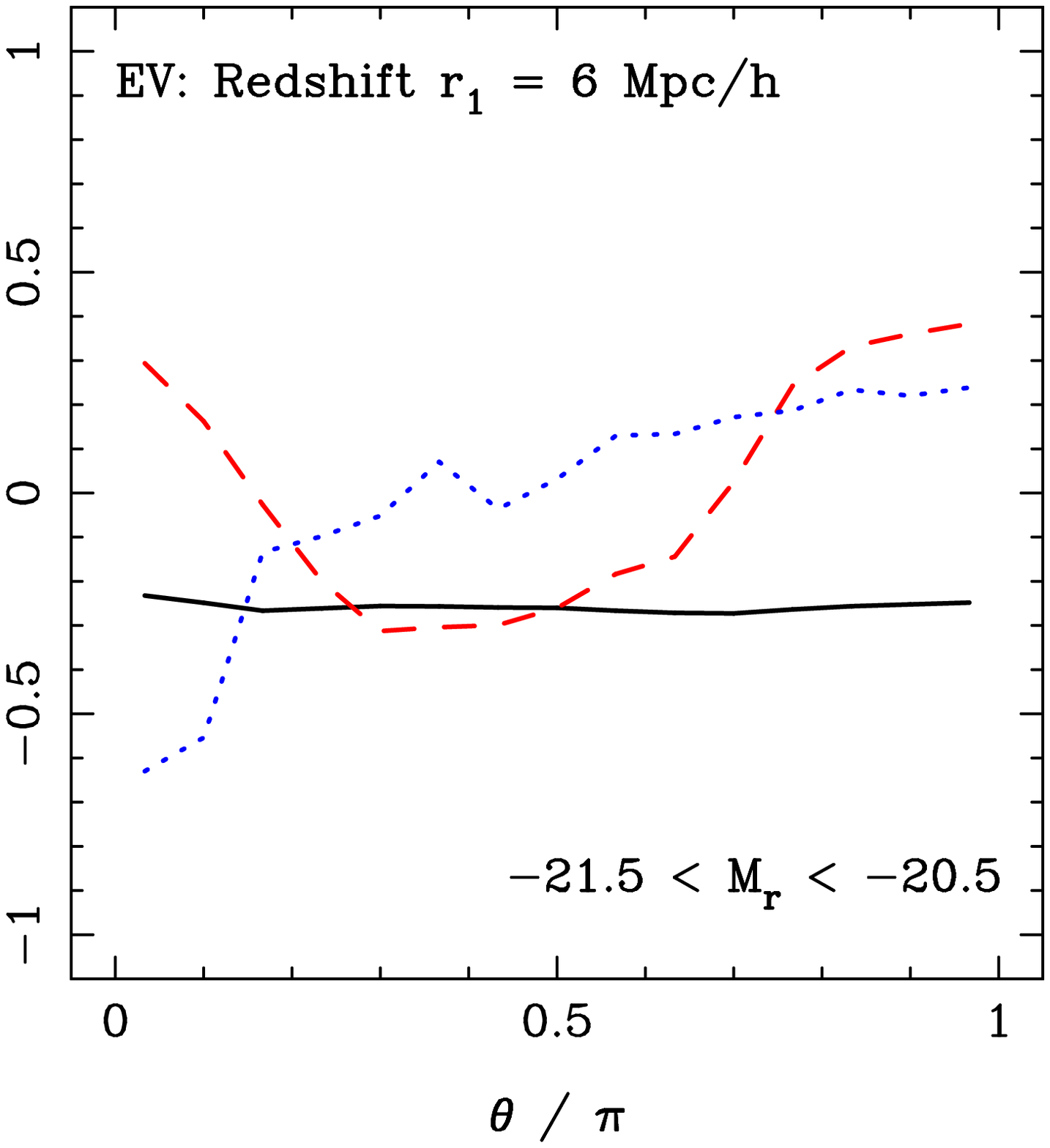}
  \includegraphics[angle=270,width=\hplotwidth]{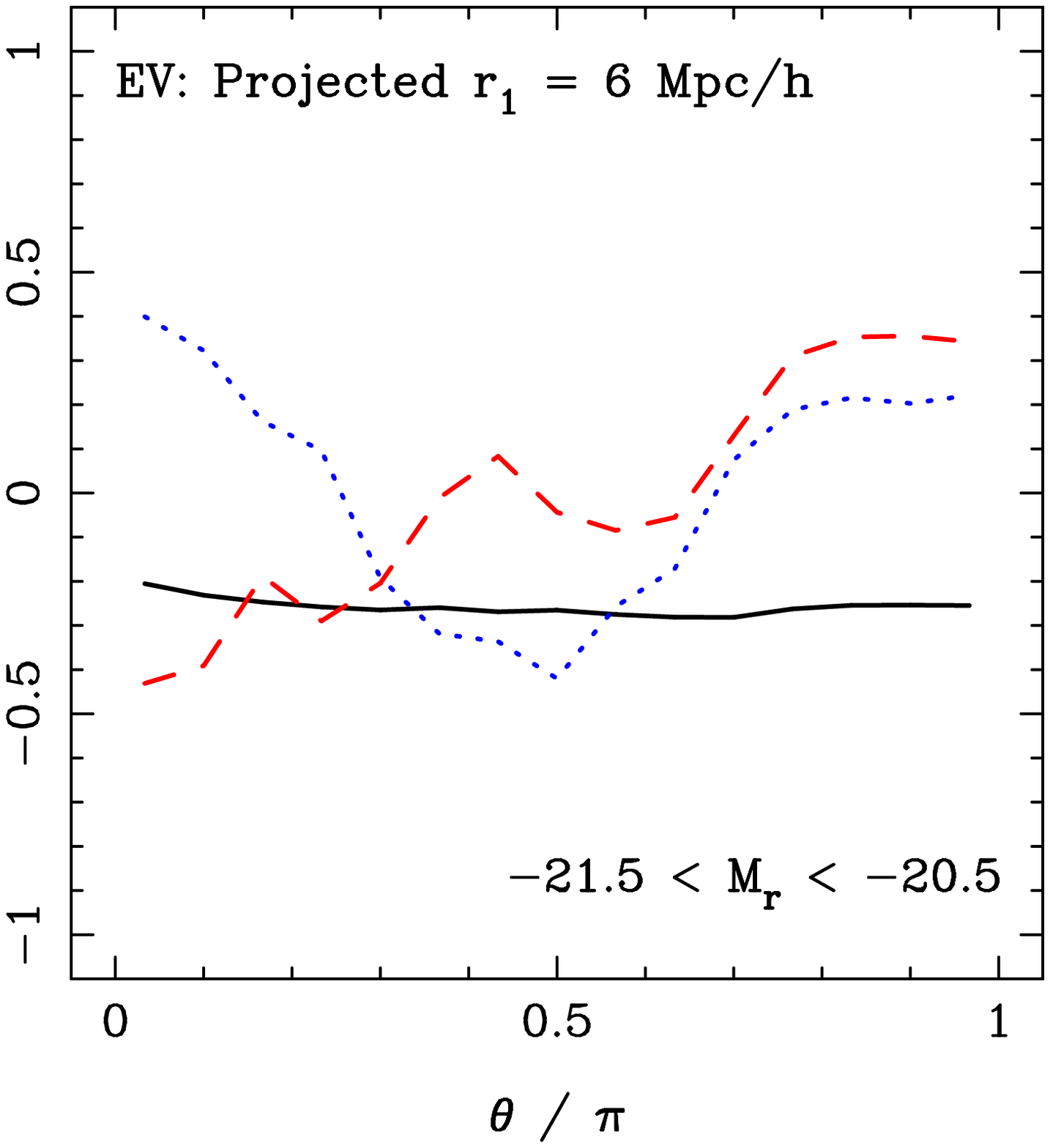}
  \includegraphics[angle=270,width=\hplotwidth]{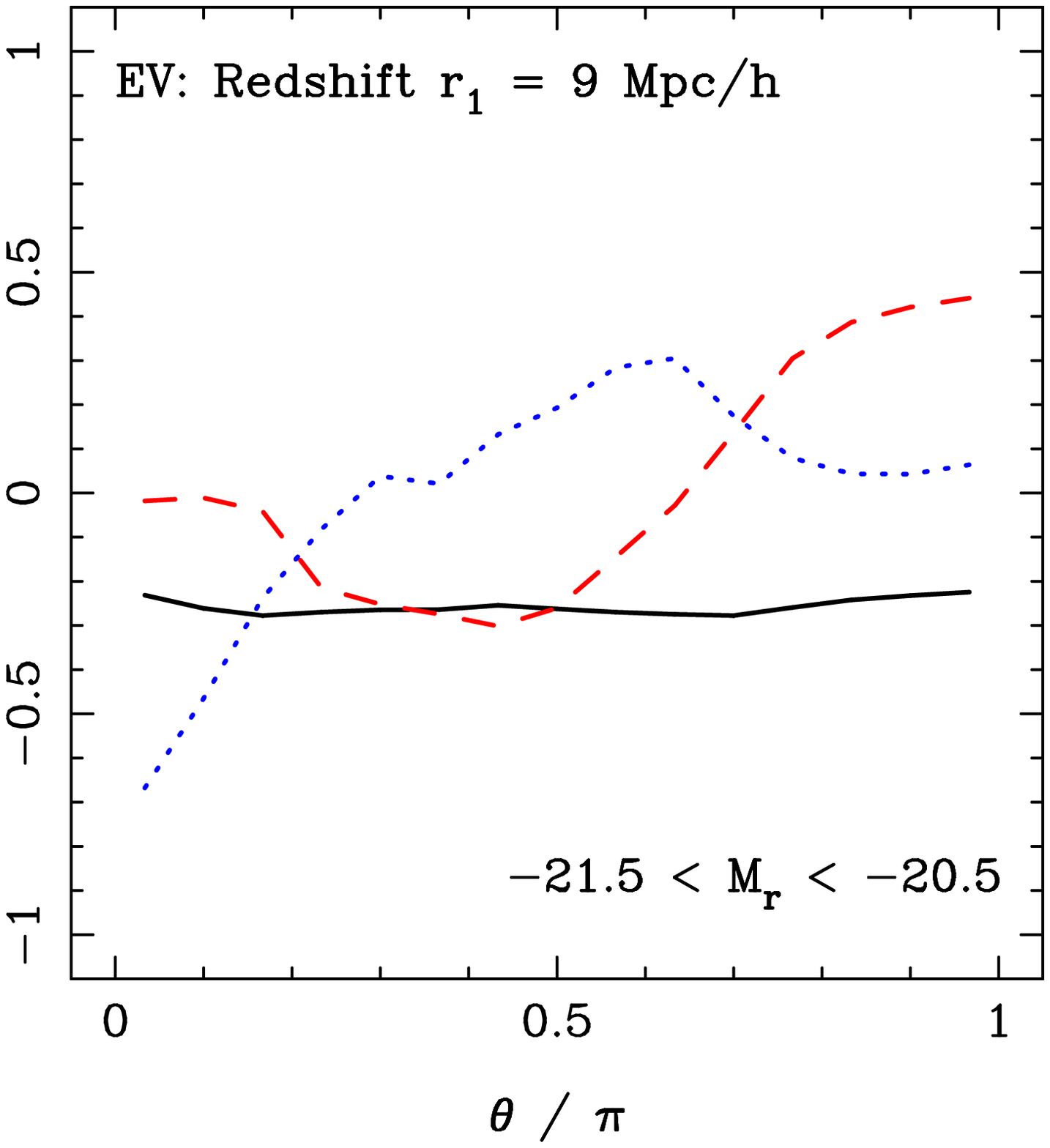}
  \includegraphics[angle=270,width=\hplotwidth]{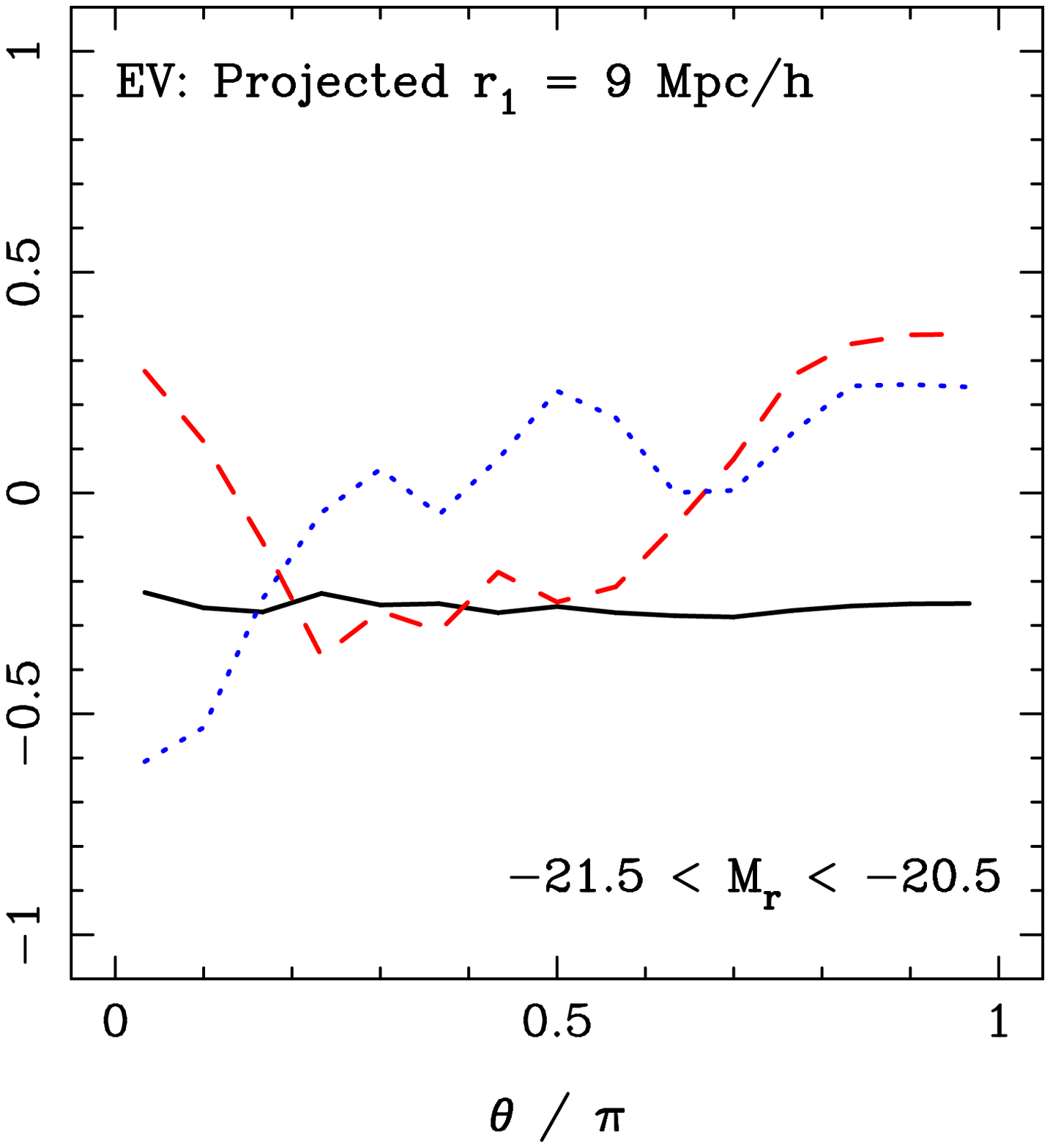}
  \caption[Top 3 Eigenvectors for $-21.5 < M_r < -20.5$]{ 
    Like Figure~\ref{f:ev3_mt21_5} but for the $-21.5 < M_r < -20.5$ galaxy sample. The top 
    three eigenvectors (EVs) chosen from the normalized covariance matrix. The sign of the 
    EV is arbitrary.  The first EV (solid black) shows equal weights for all bins.  The second (dashed red) and 
    third (dotted blue) EV display the configuration difference between perpendicular and 
    co-linear triangles as well as the scale variation as the scale of the third side increases.  
  } 
  \label{f:ev3_mb20_5} 
\end{figure}

A point that is often overlooked is that the covariance matrix itself is a measurement of 
clustering rather than simply a means of quantifying uncertainty.  
It exhibits increased sensitivity to higher 
order terms \citep[for a concise review see][]{szapudi:09}
with the covariance of the 
2PCF being leading order sensitive up to fourth order, and the 3PCF up to sixth order.

We investigate the structure of the normalized covariance matrices by examining the 
eigenvectors (EVs), or principal components, obtained by a singular value decomposition.  
The EVs are contained in the $\bm{U}$ and $\bm{V}$ matrices from \eqref{eq:svd} and 
\eqref{eq:q-eigenmodes}.  The first EV is associated with the largest singular value (SV), 
and accounts for the largest variance in the normalized covariance matrix 
(i.e. most of the observed structure); the second EV is the next largest SV and so on.  
If the covariance matrix resolves predominately ``true'' signal, the first EVs should 
characterize this structure whereas the lower ranked EVs encapsulate noise.   
While the amplitude of the EVs are not significant without the corresponding SV, they 
do represent the variation between bins in an orthogonal basis where the full covariance 
is a simple linear combination.

We show the top three EVs for the BRIGHT and LSTAR galaxy samples in 
Figures~\ref{f:ev3_mt21_5} and \ref{f:ev3_mb20_5}, respectively. 
In all cases there appear to be consistent features in the eigenmodes.  The first EV 
represents weighting all bins equally.  Typically, the second EV highlights the difference 
between ``perpendicular'' and ``co-linear'' configurations.  Finally, the third EV tracks a 
roughly monotonic change from small to large $\theta$, possibly accounting for the scale 
difference of the continually increasing third side of the triangle.  We point out that 
this structure evident in observational galaxy samples agrees well with theoretical 
predictions from simulations in \citet{GS05}.  Remember, these EVs are obtained by 
deconstructing just the \emph{normalized} covariance matrix.

We do not always see a clear separation between the second and third EVs.  As the full 
structure is a linear combination of all modes, the configuration dependence and scale 
variation effects could be combined.  The SVs of the two effects are essentially equivalent 
for our measurements, making their numerical distinction in the SVD somewhat arbitrary.  
This is not a concern, as it appears that the linear combination of these two EVs is consistent 
with our interpretation (even if they are mixed).  Less significant EVs show less coherent 
structure, consistent with noisy modes in the covariance (as we would expect).

By examining the EVs of the covariance matrices, we note structure consistent with 
measurements of the reduced 3PCF (see Figures~\ref{f:Qfit_mt21_5} and \ref{f:Qfit_mb20_5}). 
Observing this structure provides supporting evidence that we have signal dominated estimates 
of the covariance matrix.
This justifies our approach of using a combination of the most significant eigenmodes 
in a quantitative comparison to galaxy-mass bias models, as we did in \S\ref{s:bias}.

\begin{figure*}
  \centering
  \includegraphics[angle=270,width=\plotwidth]{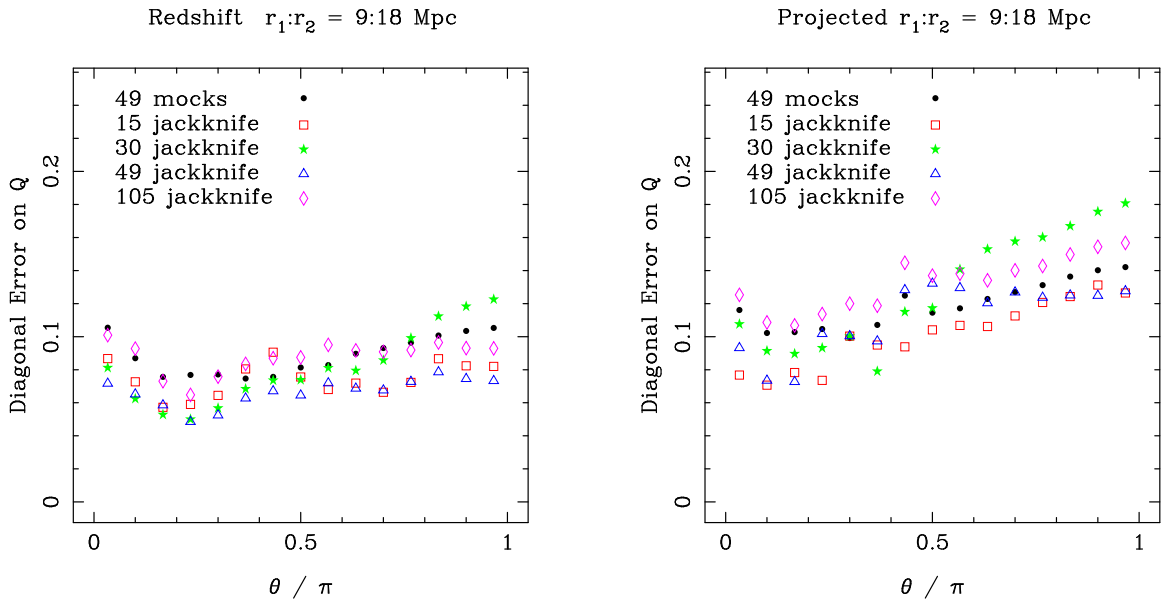}
  \caption[Absolute errors using different error estimates]{
    We compare the $1\sigma$ absolute (diagonal) errors of the reduced 3PCF obtained 
    by using different methods of estimation: independent mock catalogs or jackknife 
    resampling as denoted.  These measurements correspond to the BRIGHT galaxy sample. 
  } 
  \label{f:diagerr} 
\end{figure*}

\section{Quality of Error Estimation}
\label{s:errors}

\begin{figure*}
  \centering
  \includegraphics[angle=0,width=\plotwidth]{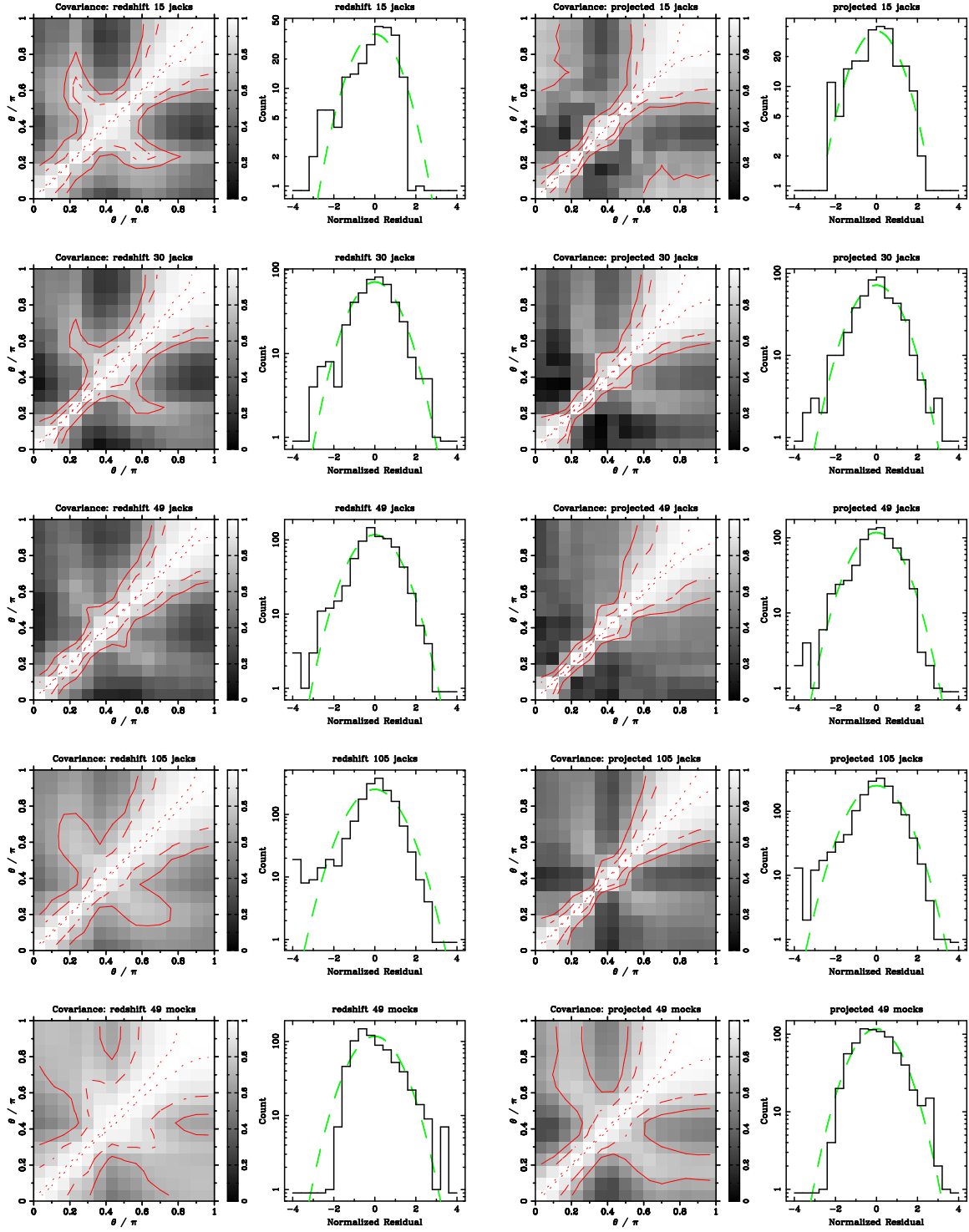}
  \caption[Comparison of covariance using different error estimates]{
    We present the normalized covariance matrices and residuals of the error estimation for large 
    triangles ($9-27 \hmpc$).  The left and right columns pertain to redshift and projected space 
    respectively.  We estimate errors using $15$, $30$, $49$, and $105$ jackknife regions and 
    compare with results from $49$ mocks from independent \nbody\ simulations.  The solid, dashed 
    and dotted contours in the normalized covariance correspond to values of 
    $0.70$, $0.85$, and $0.99$, respectively.
  } 
  \label{f:covar_cmp20} 
\end{figure*}

\begin{figure}
  \centering
  \includegraphics[angle=0,width=\hplotwidth]{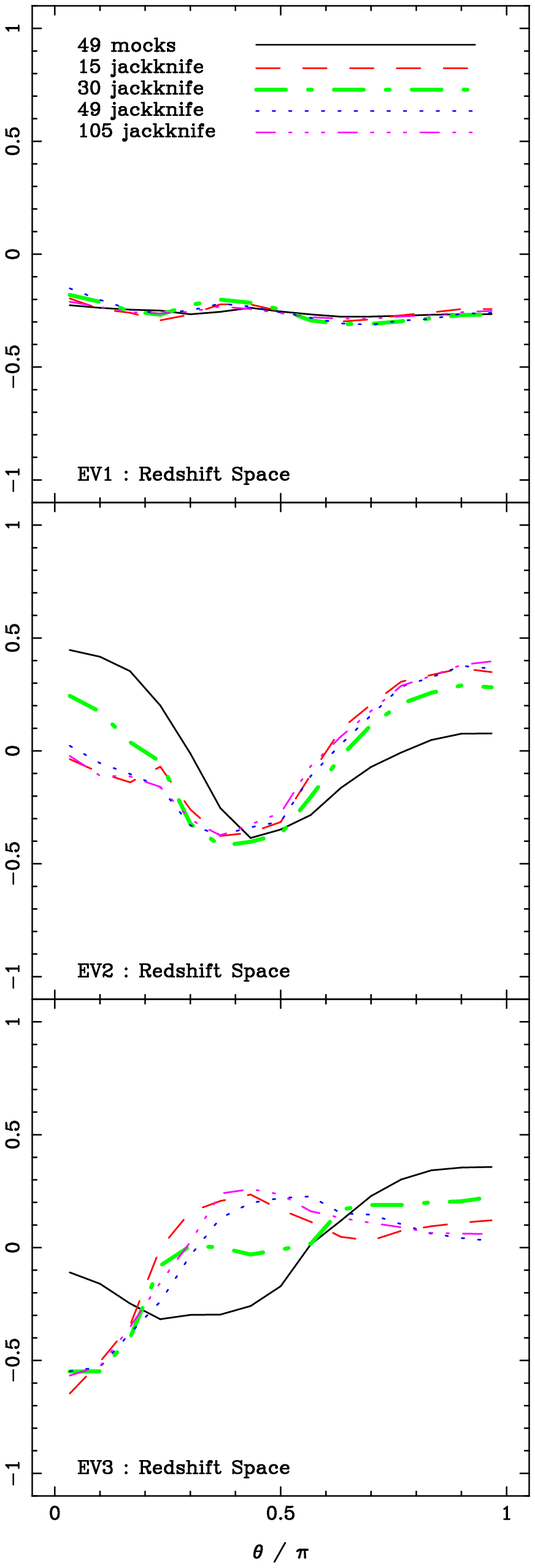}
  \qquad
  \includegraphics[angle=0,width=\hplotwidth]{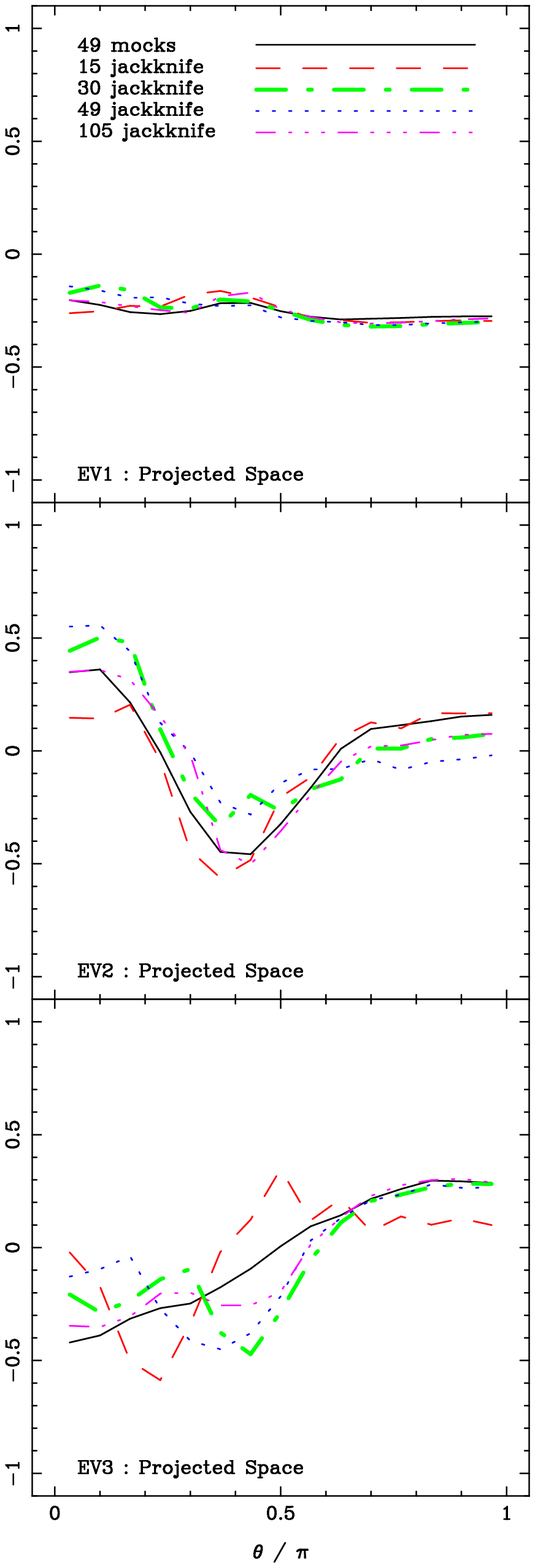}
  \caption[Top 3 Eigenvectors of normalized covariance]{
    Top three eigenvectors (EVs) chosen from the normalized covariance matrix for the different
    error estimates for $Q_z(\theta)$ and $Q_{proj}(\theta)$.  The sign of the EV is arbitrary.  
    The first EV (left panels) shows equal weights between all bins.  
    The second (middle) and third (right) EV display the configuration difference between 
    perpendicular and co-linear triangles, as well as the scale variation as the scale of the 
    third side increases.  
  } 
  \label{f:ev3} 
\end{figure}

\begin{figure}
  \centering
    \includegraphics[angle=270,width=\linewidth]{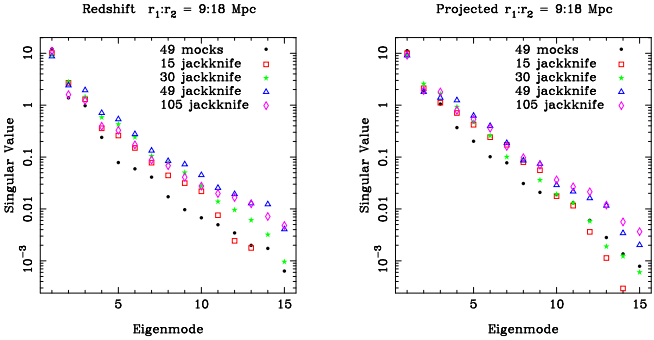}
  \caption[Singular values of the covariance matrices]{
    The singular values (SV), or eigenvalues, obtained from the singular value decomposition 
    (SVD) of the normalized covariance matrix for each of our error estimates.  
    Larger values of the SV correspond to more statistically significant eigenmodes in the 
    structure of the covariance.  
  } 
  \label{f:sv} 
\end{figure}

\begin{figure}
  \centering
    \includegraphics[angle=270,width=\linewidth]{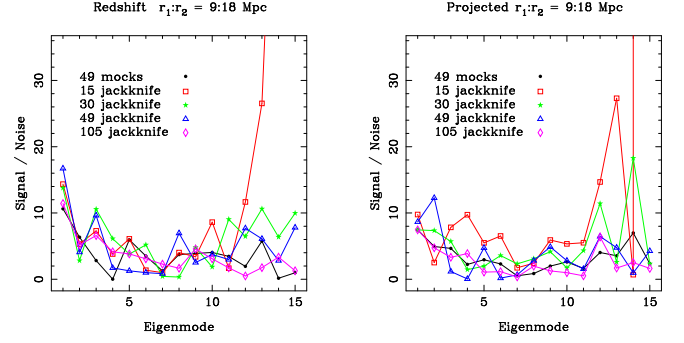}
  \caption[S/N ratio for each eigenmode]{
    The signal-to-noise ratio for each eigenmode, ordered in terms of importance.  
    The total signal-to-noise of a measurement is calculated by adding each individual 
    eigenmode in quadrature.
  } 
  \label{f:s2n} 
\end{figure}

\begin{figure}
  \centering
    \includegraphics[angle=270,width=\linewidth]{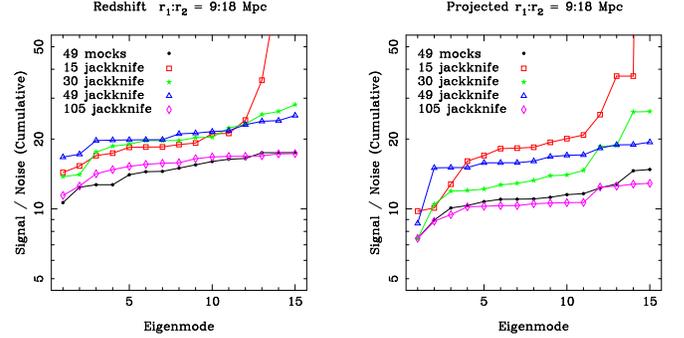}
  \caption[Cumulative S/N ratio for the eigenmodes]{
    The cumulative signal-to-noise ratio for each eigenmode, ordered in terms of importance.  
    The cumulative total is calculated by summing in quadrature the more significant modes.
  } 
  \label{f:s2ntot} 
\end{figure}

\begin{figure*}
  \centering
    \includegraphics[angle=0,width=\hplotwidth]{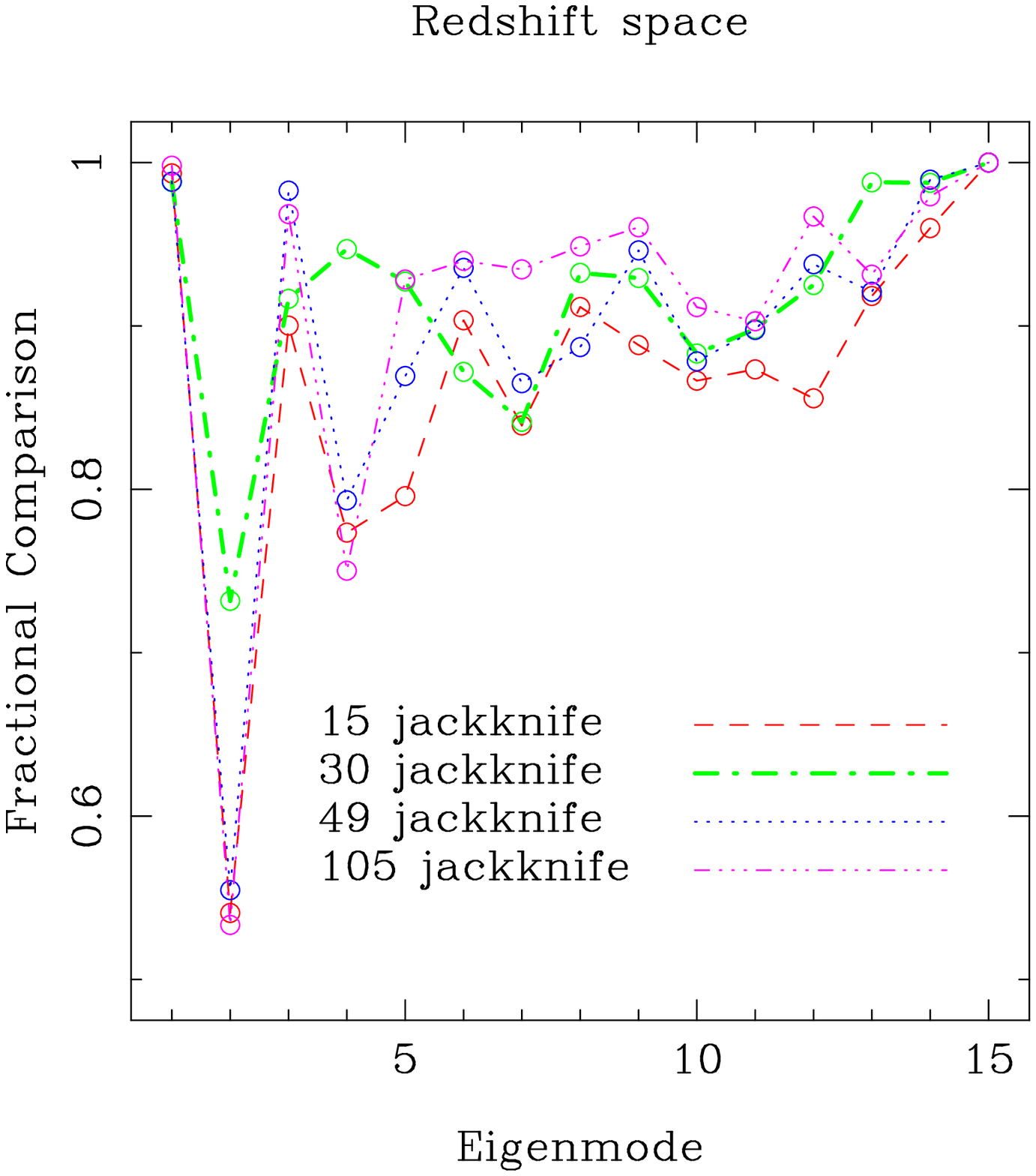}
    \qquad
    \includegraphics[angle=0,width=\hplotwidth]{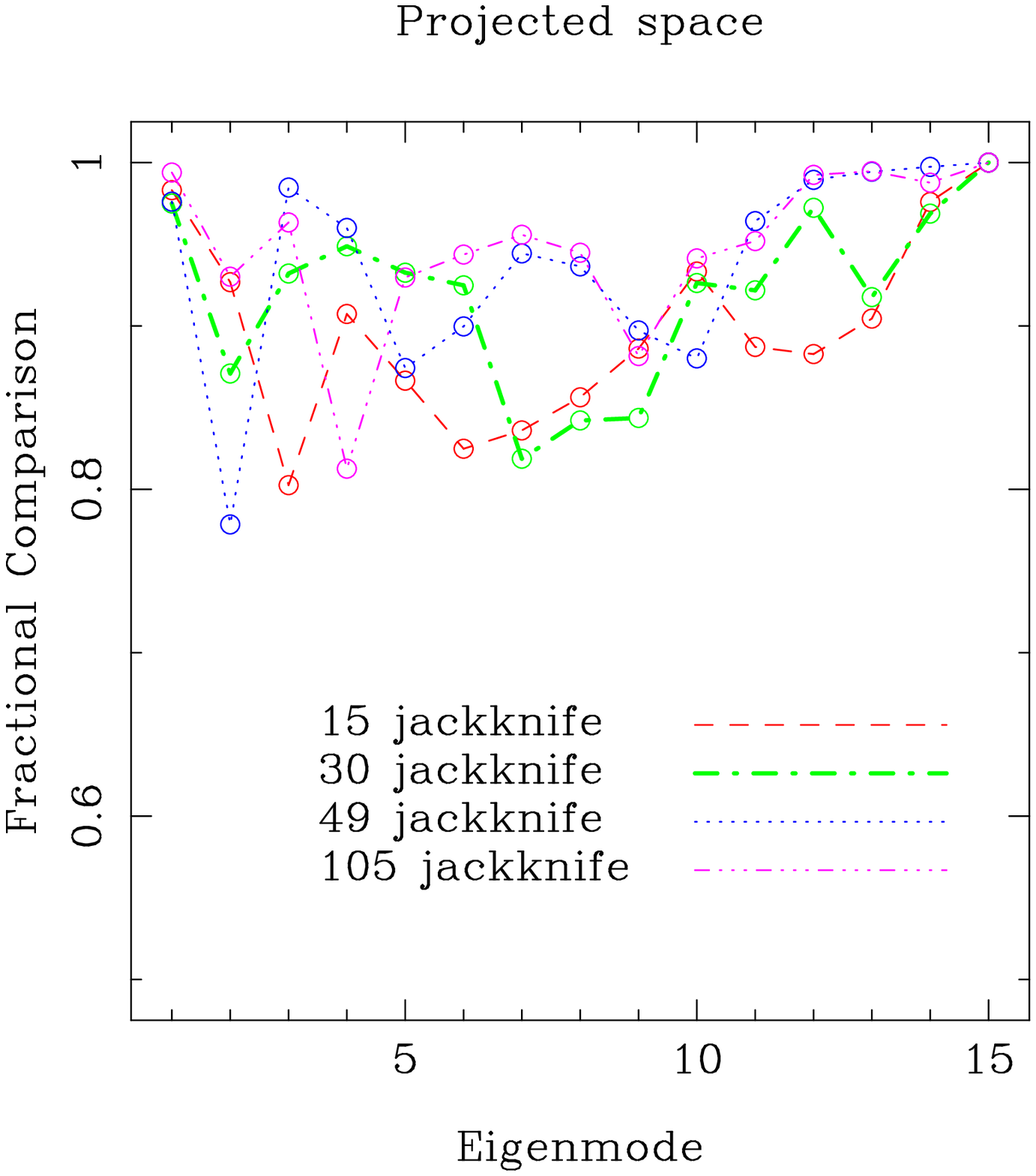}
  \caption[Eigenvector subspace comparison: jackknife vs mocks]{
    We compute the compatibility of the \emph{subspace} contained in a series of eigenvectors such 
    that the y-axis can be interpreted as the fractional "match" between the spaces the eigenvectors 
    probe.  A value of $1.0$ means no mismatch in the space they probe, and $0.0$ means no overlap
    (i.e. orthogonal).  The left panel uses the eigenvectors of the redshift space covariance
    matrix determined by the $49$ mocks as the reference value. The right plot does the same as the left, 
    but uses measurements in projected space. 
    The comparison is cumulative (eigenmode 3 means the sum of the first 3 modes). 
  } 
  \label{f:evsubspace} 
\end{figure*}

\begin{figure}
  \centering
  \includegraphics[angle=0,width=\plotwidth]{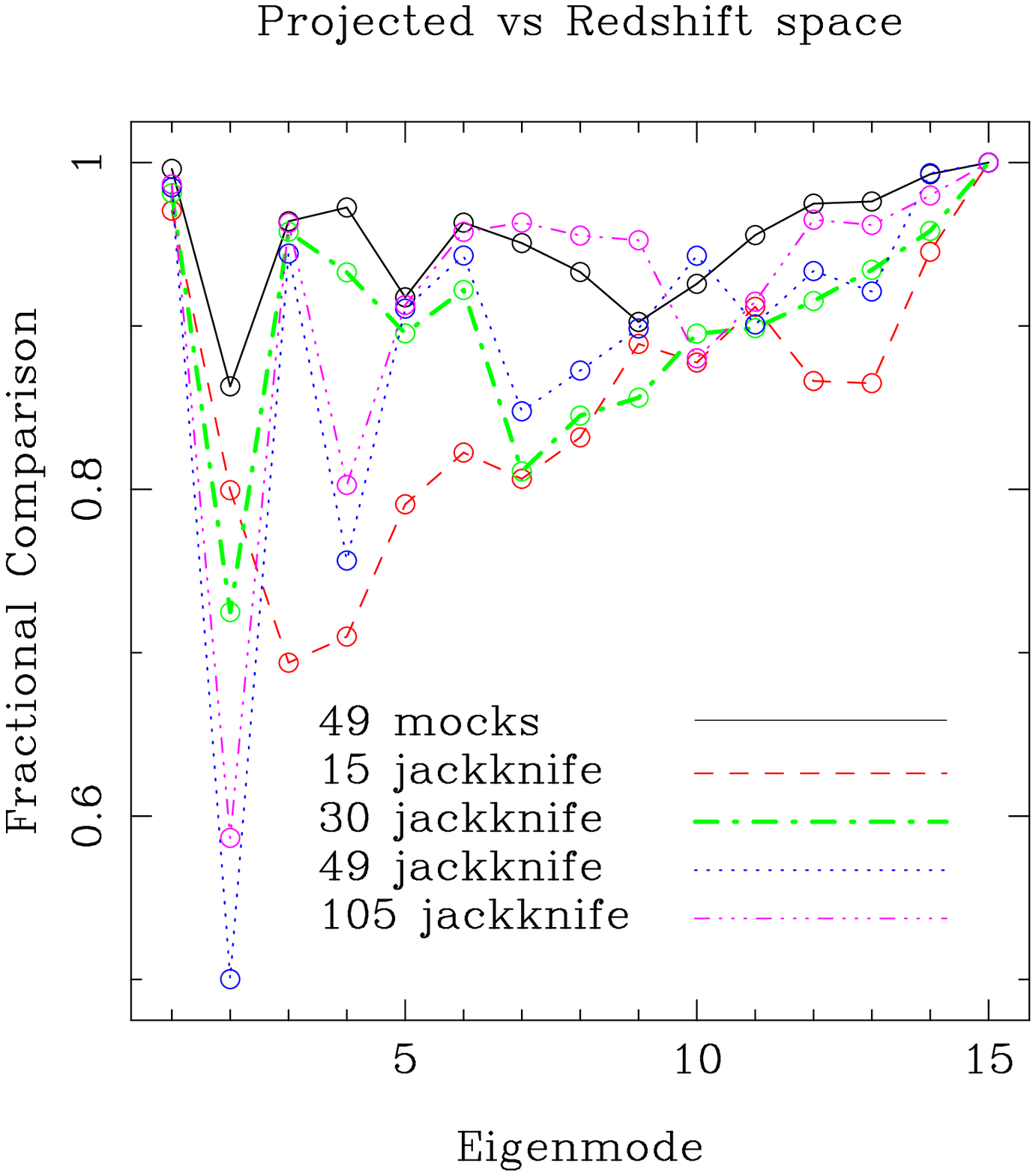}
  \caption[Eigenvector subspace comparison: redshift vs projected]{
    We use the subspace comparison of eigenvectors to estimate the difference in space probed between 
    similar numbers of eigenmodes in redshift and projected covariance matrices for each of the error 
    estimates.
  } 
  \label{f:evsubspace_ZvP} 
\end{figure}

We rely heavily on the structure of the error covariance matrix for constraints on 
galaxy-mass bias.  We noticed the observed structure in the covariance is qualitatively 
similar to clustering measurements (in \S\ref{s:ev}), but it remains unclear if this 
structure is affected by our jackknife re-sampling estimation of the covariance.  
Higher orders add complexity and increased sensitivity to systematics, even with a 
``ratio statistic'' such as the reduced 3PCF where the error sensitivity is canceled 
to first order.  We must investigate the error resolution of jackknife resampling on 
the 3PCF, as tests on the 2PCF in angular correlation functions 
\citep{scranton:02} or redshift space SDSS 
\citep{zehavi:02,zehavi:05,zehavi:10} can not be assumed to be sufficient. 

An alternative method to estimate errors uses a series of independent realizations of 
artificial galaxies, ideally created to match observational limitations such as the volume 
and geometry of the SDSS galaxy samples.  We created 49 independent galaxy mock catalogs 
based on independent \nbody\ simulations that have appropriate resolution to match the 
BRIGHT SDSS sample ($M_r < -21.5$).  We use these independent mocks to estimate 
errors and compare with those obtained from jackknife re-sampling of \emph{the data}.  
This exercise should provide an idea how effective jackknife re-sampling is for resolving the errors 
on the 3PCF. The BRIGHT galaxy sample has the lowest number density with the least number 
of galaxies over the largest volume.  To help protect against undersampled measurements due 
to low bin counts \citep[see discussion in ][]{mcbride:10}, we restrict the comparison to 
the configuration dependence of the larger triangles ($r_1 = 9 \hmpc$ sides).  

We estimate the covariance matrix for the BRIGHT sample using different numbers of 
jackknife samples on \emph{the data}, specifically using $15$, $30$, $49$ and $105$ 
jackknife regions.  Again, these jackknife sample are created from the observational data
and not from the mock galaxy catalogs, where each jackknife region is selected to maintain 
equal unmasked area \citep[same method as detailed in ][]{mcbride:10}. 
Since we measure $15$ bins for $Q(\theta)$, we require at least $15$ 
jackknife samples to prevent a singular covariance matrix.  We use twice this number 
($30$) and then use the same number as the number of mocks ($49$).  The final value
corresponds to the number of unique elements in the symmetric covariance matrix: 
$15(15-1)/2 = 105$.  We caution that as we increase the number of jackknife samples, we 
decrease the respective volume of each jackknife region which might subtly bias jackknife 
estimates (e.g. underestimate cosmic variance).

First, we investigate the magnitude of the absolute (diagonal) errors of the reduced 3PCF.
Since we use normalized covariance matrices, differences in the absolute errors might not 
be noticeable in the covariance structure.  The $1\sigma$ absolute errors are shown in 
Figure~\ref{f:diagerr}.  We see little difference between any of the methods, and the 
uncertainty typically ranges between $0.1$ and $0.15$.  

For each of our methods, we estimate the normalized covariance in both redshift and projected 
space, as depicted in Figure~\ref{f:covar_cmp20}, and include the distribution of 
residuals.  We note that jackknife re-sampling methods appear to underestimate the 
correlation in all cases, but the general structure looks comparable.  
More samples generally produce a smoother, and more correlated, covariance matrix.  However, 
not even the $105$ jackknife sample estimate reproduces the correlation in $49$ mocks.  
We consider the distribution of residuals an important metric in evaluating the 
reliability of the resulting covariance, which we include in Figure~\ref{f:covar_cmp20}.  
Ideally, the covariance matrix accounts for all ``connections'' between bins only if the 
residuals are reasonably Gaussian.  We notice a skew in several of the jackknife re-samplings, 
with a tail extending to lower values.  As discussed in \citet{mcbride:10}, this is a 
consequence of cosmic variance within jackknife samples.  A few rare structures affect the 
3PCF; when they are excluded by an a jackknife region the $Q(\theta)$ of the entire 
sample drops.  The mock estimate shows a slight skew in the positive direction from the 
same effect.  In mocks, when a rare structure exists in the probed volume then the 3PCF 
rises producing a rare high measurement.

The eigenmode analysis we utilize relies on signal being the dominant contribution to the 
structure of the covariance matrix (as opposed to noise).  Noise is commonly expected to 
be an independent or \emph{diagonal} contribution.  Similar to \S\ref{s:ev}, we examine the 
eigenvectors (EVs) of the covariance matrix to provide insight into the structure.  
By using the singular value decomposition (SVD), the eigenvectors are 
ordered by largest to least amount of variance explained in the covariance matrix. 

The first three EVs are shown in Figure~\ref{f:ev3} for both redshift and projected space.  
Similar structure appears in each of them, which we interpret as follows.
The first EV represents the general measurement, 
with all eigenmodes equally weighted.  The second EV shows the difference between 
``collapsed'' and ``perpendicular'' configurations.   Finally, the third EV
represents a scale dependence as the third side of the triplet ranges between 
$9 \hmpc$ at $\theta \sim 0$ to $27 \hmpc$ at $\theta \sim \pi$.  
In some of the estimates, the shapes of the second and third EVs appear either combined 
or transposed.  Since the full measurement is a linear combination of all EVs, this lack 
of separation makes sense.  In these cases, the statistical significance the two EVs 
remain similar.  This interpretation of the structure follows the analysis by 
\citet{GS05} for \nbody\ simulations.  The less significant eigenvectors 
(which we do not show) appear random, with the lowest being contributions from noise or 
numerical instabilities.  We identify the significance of the eigenmodes by inspecting the 
singular values (SVs) shown in Figure~\ref{f:sv}.  The SV can be understood as an 
``importance weighting'' of each eigenmode, and the figure shows a rapid decline of
significance for each eigenmode.  
The first three eigenvectors cumulatively account for over $99.9\%$ of the variance in 
the normalized covariance matrix.  

The signal-to-noise ratio ($S/N$) of each eigenmode is shown in Figure~\ref{f:s2n}, as 
calculated by \eqref{eq:s2n}.  The mocks in both redshift and projected space depict a slow decline 
in $S/N$ over the first few eigenmodes, supportive or our interpretation of relative significance.  
This trend is not as clear in the jackknife estimates for redshift space, although it 
appears consistent in projected space.  We see the first half of the modes appear resolved, 
with well behaved $S/N$.  For the least significant eigenmodes, the noisier error estimates 
using fewest jackknife samples show unrealistically high $S/N$ ratios (especially in the 
case of $15$ jackknife regions).  The total $S/N$ would increase dramatically and artificially
if we included these noise dominated modes.  In these cases, using the full covariance 
(i.e. including all modes) would be a mistake.  To make the point clearer, we examine the 
cumulative $S/N$ ratio in Figure~\ref{f:s2ntot} where we identify rapid upturns in the 
total $S/N$ as an artificial consequence of noise.  Several curves in Figure~\ref{f:s2ntot} do 
not appear problematic with this test, and show steady behavior across all modes.  
The amplitude of the $S/N$ ratio between $49$ mocks and the $105$ jackknife samples 
show consistency, but the $S/N$ ratio does not appear to be a monotonic change 
with the number of jackknife samples which suggests a complex relationship between the 
best $S/N$ and an optimal number of jackknife regions. 

We can compare the \emph{subspace} that a set of eigenvectors probe between two error 
estimates.  The formalism is the same as discussed in \citet[ see section 4]{yip:04}, which 
results in a fractional ``compatibility'' between a collection of eigenvectors.  Intuitively, 
this is the matrix equivalent of the vector dot product, where two orthogonal 
unit vectors would have a vector subspace of $0$ (no compatibility) and two identical unit 
vectors would result in $1$.  We use the covariance of the $49$ mocks as ``truth'', and 
test the fractional compatibility of the jackknife estimates for covariance 
in $Q_z(\theta)$ and $Q_{proj}(\theta)$ shown in Figure~\ref{f:evsubspace}.  
When all the eigenmodes are considered, the \emph{subspace} becomes the full space 
and the comparison yields unity by construction.  We notice the projected measurements 
never appear more discrepant than $75\%$.  After the first few eigenmodes, redshift space 
shows a similar agreement.  With the exception of the $15$ jackknife sample estimate, 
the 3 eigenmode mark appears $90\%$ compatible or better in all cases.  This quantifies 
our argument of the top three EVs in Figure~\ref{f:ev3}, where the second and third 
eigenvectors appear different (predominantly in redshift space), but their linear 
combination remains consistent with each other.  Remember, this comparison only 
considers the compatibility of the \emph{direction} of each eigenvector, and not their 
relative strengths (i.e. SVs).

We evaluate the subspace compatibility on the normalized covariance matrix between redshift 
and projected space estimates.  For each method, we show the fractional comparison 
in Figure~\ref{f:evsubspace_ZvP}.  The mocks estimates show the most compatibility across 
all eigenmodes, where showing agreement at $\sim 85\%$ or better.  With all estimates, 
we find that the combination of the first three eigenmodes remains a compatible subspace, 
above $90\%$, if we again exempt the $15$-jackknife sample estimate (which shows less than 
$70\%$ compatibility).

We caution that the resolution of errors and the choice of binning scheme relate in a 
non-trivial manner, which is discussed in additional detail by \citet{mcbride:10}.  
We chose ``large'' bins (fiducial scheme with $f=0.25$) to ensure a smooth, signal dominant 
structure in the covariance matrix.  Overall, this error comparison supports our 
claim that accurate results can still be obtained even with less-than-optimal error 
estimation such as jackknife re-sampling.  


\section{Discussion}
\label{s:disc}

We utilize the the configuration dependence of the 3PCF in redshift and projected space
to constrain galaxy-mass bias parameters in the local bias model.  We find that galaxies 
are biased tracers of mass, with brighter galaxies corresponding to increased 
bias.  These results are consistent with detailed analysis of SDSS galaxies from the 
2PCF \citep{zehavi:05,zehavi:10} which quantifies how bias increases clustering for 
brighter galaxy samples.  Our results indicate that a linear bias model yields reasonable 
approximations to the observations, in agreement with \citet{hikage:05}. However, 
a non-linear bias model produces slightly better agreement, and yields lower reduced 
chi-square values ($\chi^2_\nu$ in Tables~\ref{t:bias} and \ref{t:linbias}).  
We notice a strong correlation between linear and quadratic bias, as expected from inspection of 
\eqref{eq:biasQ}, and consistent with measurements of SDSS galaxies using the bispectrum \citep{nishimichi:07}.  
We find that our redshift space measurements predict significantly negative quadratic 
bias with a linear bias near one.  This effect was seen in a similar analysis conducted 
on 2dFGRS galaxies \citep{gaztanaga:05}.  Interestingly, we find projected measurements 
suggest a larger linear bias with near zero quadratic bias for the same samples, 
suggesting a possible systematic effect from redshift distortions in this simple bias 
model.

We examined the relative bias in \S\ref{ss:brel}.  We find supporting evidence that the 
brighter galaxy sample is a \emph{more biased} realization using both the 2PCF and 3PCF, 
consistent with other analyses of SDSS data \citep{zehavi:02,zehavi:05,zehavi:10}.  
Relative bias provides a consistency check on the ``absolute'' galaxy-mass bias parameters 
we constrain, suggesting a combination of linear and quadratic bias terms are consistent 
with observations.  However, the relative bias of the 2PCF suggests that our two parameter 
bias model fits underpredict the value of linear bias necessary to explain the 
observations.  Again, we see a hint that constraints from projected measurements appear 
to be less affected -- although we caution that this trend has weak statistical 
significance given the larger uncertainties in projected space.

We obtain reasonable projections for $\sigma_8$ by using our linear bias values from fits 
on $Q(\theta)$ in conjunction with the 2PCF.
We estimate the values of $\sigma_8$ to be between $0.83$ and $1.13$ based on the BRIGHT 
($M_r < -21.5$) and LSTAR ($-21.5 < M_r < -20.5$) galaxy samples.  The values we obtain 
are contingent on a specific model of mass clustering, where we have chosen to use \nbody\ 
simulations (specifically the Hubble Volume \lcdm\ results), and redshift distortions 
(which we include through velocity information to distort particle positions in the HV 
simulation).  For comparison, constraints of $\sigma_8$ from a joint analysis of the 
cosmic microwave background (CMB), supernova data (SN) and baryon acoustic oscillations 
(BAO) find $\sigma_8 = 0.82$ \citep{komatsu:09}.
Our lower values are in good agreement with these constraints.  Our high end values 
appear too large, but our results are in reasonable agreement with an analysis of a 
related statistic, the monopole moment of the 3PCF, where they find best fit $\sigma_8$ 
values between $0.9$ and $1.07$ \citep[see Table~3 in][]{pan:05} using 2dFGRS galaxies 
\citep{2dFGRS}.  Although the value of $\sigma_8$ we obtain is comparable with results 
from 2dFGRS, the specific bias values will not be, as the 2dF targets a different galaxy 
selection than our SDSS samples.  If we underestimate the value of linear bias, effectively $B$ here, 
\eqref{eq:bias2pt_s8} shows that the implied value of $\sigma_8$ will be overestimated.  
This might explain the larger values of our estimates in comparison to WMAP analyses.  
Our projections for $\sigma_8$ use clustering measurements between $9-27 \; \hmpc$ and 
exploit only the configuration dependence of $Q(\theta)$. This is a much smaller slice of 
data that is significantly different than either the monopole measurement (which utilizes a 
larger range of scales without configuration dependence) or WMAP results (that combines a 
immense amount of data from both CMB and LSS analyses).  We do not intend this analysis to 
complete with these constraints, but rather to help illuminate the role of galaxy-mass 
bias in future constraints of $\sigma_8$ using the 3PCF.

Understanding the properties of measurement errors and the impact of empirical methods of 
estimating the covariance is a critical component necessary for quantitative constraints.  
Recent results have done comparisons on lower order statistics, such as the work by 
\citet{norberg:09}.  We compared several properties of 3PCF covariance matrices estimated 
from jackknife re-sampling to those constructed from many realizations of independent 
galaxy mock catalogs.  While we noted some concerning discrepancies, we found these 
typically affected only the least significant eigenmodes.  We found many similarities 
between the covariance estimates, including physical descriptions for the first three eigenmodes 
which account for an overwhelming majority of the variance.  
We established the need to trim noisy, unresolved 
modes from the covariance.  When trimmed, and the eigenmode analysis is properly utilized, 
we noted only a few significant differences, mostly in the case of $15$ jackknife samples.  
We conclude that our use of $30$ jackknife samples does not significantly affect our 
analysis. 

\section{Summary}
\label{s:summary}

We analyze measurements of the configuration dependence of reduced 3PCF for two SDSS 
galaxy samples that were first presented in \citet{mcbride:10}.  In both redshift and 
projected space, we characterize the galaxy clustering differences with those predicted by 
the non-linear mass evolution in the \lcdm\ Hubble Volume simulation.
Here, we summarize our main results: 

\begin{itemize} 
  \item We demonstrate that brighter galaxies remain a more biased tracer of the mass field
    by constraining the linear and quadratic galaxy-mass bias parameters using a maximum 
    likelihood analysis on scales between $6$ and $27 \hmpc$.  Conservatively using scales 
    above $9 \hmpc$, the BRIGHT sample is biased at greater than $2\sigma$ and the fainter 
    LSTAR shows no significant bias, in generally agreement with expectations from 
    previous analyses of SDSS galaxies \citep{zehavi:05,zehavi:10}.  The bias parameters 
    and their significance are summarized in Table~\ref{t:bias}.

  \item We resolve the degeneracy between the linear and quadratic bias terms, which helps 
    to explain the weak luminosity dependence observed in the reduced 3PCF. 

  \item We find a linear bias model appears sufficient to explain the measurements of 
    the 3PCF by re-fitting the linear bias while constraining the quadratic bias at
    zero (results reported in Table~\ref{t:linbias}).  However, we find the two parameter 
    fit is preferred in our likelihood analysis, as it yields a lower chi-square in the 
    best fit value.

  \item The relative bias between samples of different luminosities (which is independent of 
    the mass predictions), as well as the cosmological implications for values of $\sigma_8$ , 
    show general consistency with previous analyses.  Inspection of our results suggest 
    that the linear bias values obtained without a quadratic bias term are preferred.  
    This suggests that two-parameter bias constraints might underpredict the linear bias. 

  \item We decompose the structure of the normalized covariance matrix as an alternative 
    view into clustering properties of our samples. The eigenvectors of the first three 
    dominant modes show coherent structure consistent with variations seen in the 
    $Q(\theta)$ measurements, supporting our claim that the covariance is signal 
    dominated and sufficiently resolved.  

  \item We find that jackknife re-sampling methods cannot reproduce the correlation seen 
    in the a 3PCF covariance matrix estimated from many realizations of mock galaxy catalogs.
    By performing a detailed comparison of the properties and structure of the errors, we 
    identify that noisy, unresolved modes introduce significant discrepancies.  We find that 
    using an eigenmode analysis can mitigate the differences and conclude that our analysis 
    should not be significantly affected by less-than-ideal methods of error estimation.

  \item Comparing results between redshift space and projected measurements implies a 
    potential systematic bias on values from the redshift space analysis when 
    scales below $9 \hmpc$ are included, which have been utilized in other comparable analyses. 
    Since the small scale measurements contain more constraining power than larger scales, 
    they drive the likelihood analysis even when larger scales are considered. 

  \item On scales above $9 \hmpc$, the statistical significance of constraints from 
    redshift space analyses appear stronger than those found in analyses of projected measurements. 
    We attribute this result to the increased uncertainties of the projected 3PCF, which mixes 
    in larger scales (with larger errors) due to the line-of-sight projection.  When 
    considered with the results of \citet{mcbride:10}, which finds the projected 3PCF 
    recovers configuration dependence at small scales lost in redshift space, a 
    combination of redshift space analysis at large scales and projected measurements at
    small scales would form a nice complement in future analyses. 
    
\end{itemize} 

\acknowledgments 

We thank many in the SDSS collaboration, where active discussion helped to refine this work.  
We would like to specifically acknowledge valuable input from Istv\'an Szapudi, 
David H. Weinberg, Zheng Zheng, Robert Nichol, Robert E. Smith, Andrew Zentner, and 
the detailed discussions on error estimates with Idit Zehavi.

We thank August Evrard and J\"{o}rg Colberg for kindly providing data and assistance with 
the Hubble Volume (HV) simulation.  The HV simulation was carried out by the Virgo 
Supercomputing Consortium using computers based at the Computing Centre of the Max-Planck 
Society in Garching and at the Edinburgh parallel Computing Centre.  

J.~G. and the development of \ntropy\ was funded by NASA Advanced Information Systems Research Program grant NNG05GA60G.
A.~J.~C. acknowledges partial support from DOE grant DE-SC0002607, NSF grant AST 0709394, and parallel application development under NSF IIS-0844580.

%
%
%
%

This research was supported in part by the National Science Foundation through TeraGrid 
resources provided by NCSA (Mercury) and the PSC (BigBen) under grant numbers 
TG-AST060027N and TG-AST060028N.

Funding for the SDSS and SDSS-II has been provided by the Alfred P. Sloan Foundation, the 
Participating Institutions, the National Science Foundation, the U.S. Department of 
Energy, the National Aeronautics and Space Administration, the Japanese Monbukagakusho, 
the Max Planck Society, and the Higher Education Funding Council for England. The SDSS Web 
Site is \texttt{http://www.sdss.org/}.

The SDSS is managed by the Astrophysical Research Consortium for the Participating 
Institutions. The Participating Institutions are the American Museum of Natural History, 
Astrophysical Institute Potsdam, University of Basel, University of Cambridge, Case 
Western Reserve University, University of Chicago, Drexel University, Fermilab, the 
Institute for Advanced Study, the Japan Participation Group, Johns Hopkins University, the 
Joint Institute for Nuclear Astrophysics, the Kavli Institute for Particle Astrophysics 
and Cosmology, the Korean Scientist Group, the Chinese Academy of Sciences (LAMOST), Los 
Alamos National Laboratory, the Max-Planck-Institute for Astronomy (MPIA), the 
Max-Planck-Institute for Astrophysics (MPA), New Mexico State University, Ohio State 
University, University of Pittsburgh, University of Portsmouth, Princeton University, the 
United States Naval Observatory, and the University of Washington.


\begin{appendix}

\section{Effects of Binning}
\label{s:binning} 

\begin{figure*}
  \centering
  \includegraphics[angle=0,width=\hplotwidth]{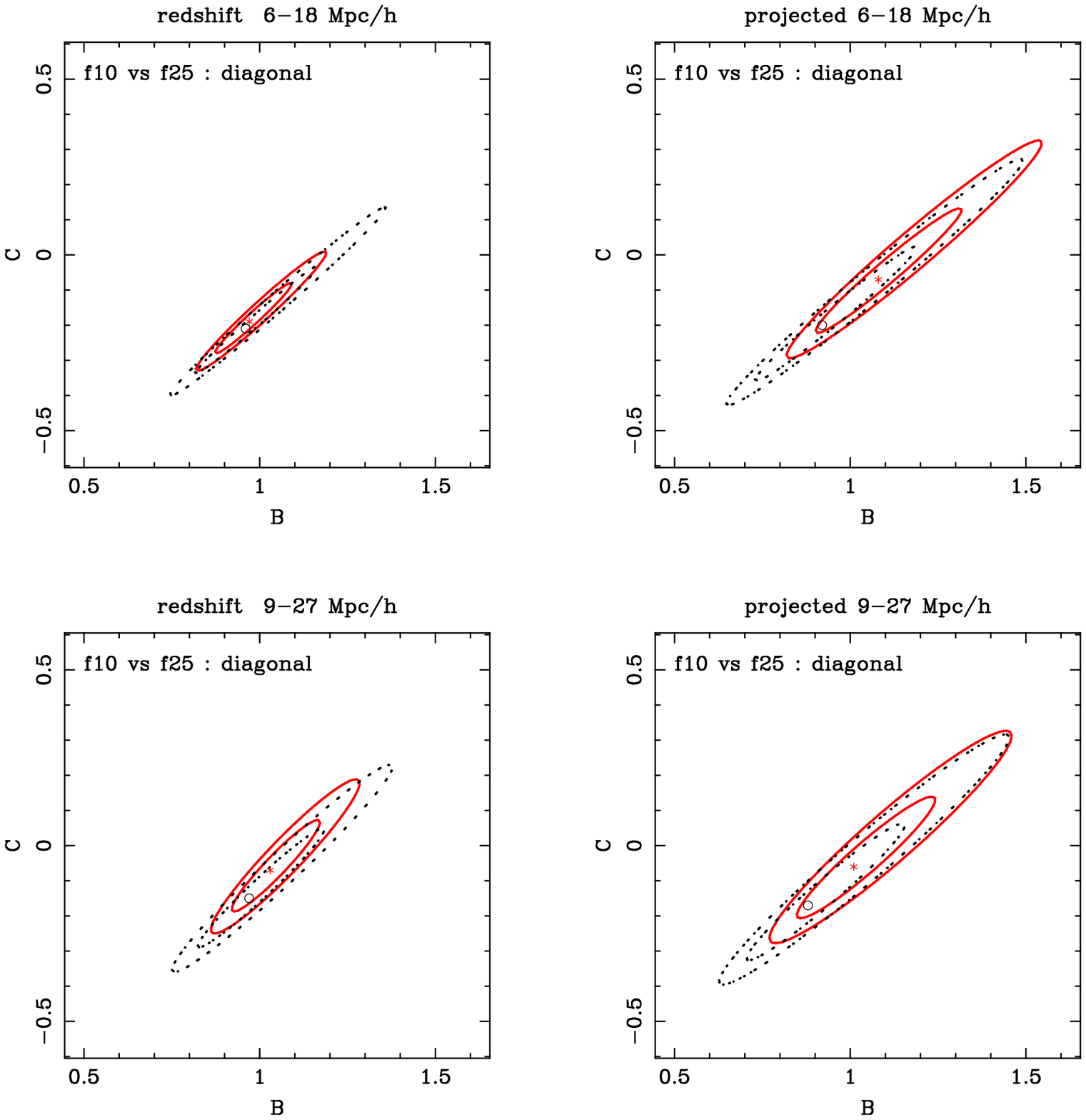}
  \qquad
  \includegraphics[angle=0,width=\hplotwidth]{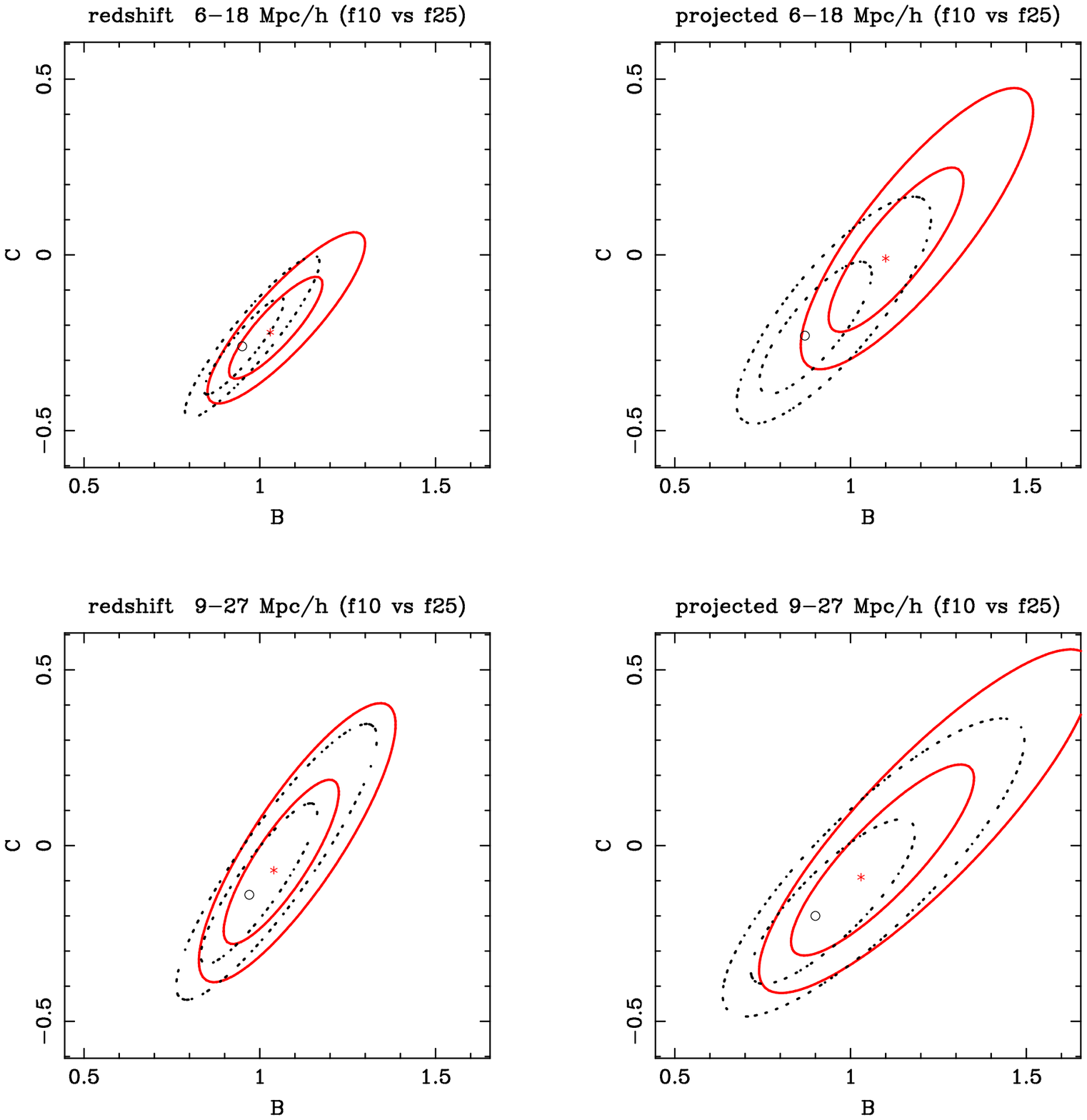}
  \caption[Differences in Galaxy-Mass bias due to binning]{
    Analogous to the Galaxy-Mass bias constraints in \S\ref{s:bias}, we show the constraints on
    $B$ and $C$ using the same data for our fiducial binning scheme with fractional bin-size of
    $f=0.1$ (solid red contours) and larger $f=0.25$ (block dotted).  
    On the left, we neglect any overlap as well as the the covariance and assume independent 
    diagonal errors while and use the full 15 bins.  
    On the right, we utilize the full covariance and only fit the dominant modes in an 
    eigenmode analysis.  
    The contours correspond to the $1\sigma$ and $2\sigma$ confidence levels from the 
    $\Delta\chi^2$ surface.  We use the LSTAR galaxy sample.
  } 
  \label{f:bin_bias} 
\end{figure*}

As an example of the effects of bin-size on galaxy-mass bias constraints, we re-analyze 
the LSTAR galaxy sample ($-21.5 < M_r < -20.5$) using the two fractional bin-widths: 
$f=0.1$ and $f=0.25$.  First, we ignore the structure of the covariance matrix and show 
constraints using all the bins assuming perfect independence shown in left panel of 
Figure~\ref{f:bin_bias}.  While unphysical, this illustration allows one to probe the 
effect of shape differences in the 3PCF measurements without considering the resolution of 
the covariance matrix. Since larger bins smooth the configuration dependence, we expect a 
larger degeneracy between $B$ and $C$, which is apparent.  We see the best fit values 
(symbols) stay within the respective $1\sigma$ contours, but just barely.  Remember, this 
approach uses all modes and the exact same input data, suggesting that binning can result in a 
$1\sigma$ systematic bias.

For the right panel of Figure~\ref{f:bin_bias}, we consider the full covariance as well as improvements obtained by
using the eigenmode analysis in the galaxy-mass bias constraints.  First, we notice the error
contours appear less stretched, in accord with our expectations of using a non-diagonal covariance 
matrix. In most cases (excepting $Q_z$ for $r_1 = 6 \hmpc$), the area of the contours appear of equal 
size or even decreased for the larger $f=0.25$ measurements in contrast to the diagonal case.  
This makes sense, as the lower variance measurements of $f=0.25$ appear better resolved as long
as there are enough remaining modes to constrain two parameters.  The best fit values appear
discrepant, especially at the lower scales ($6-18 \; \hmpc$) where they disagree at more than a 
$1\sigma$ significance.  While this causes some concern, it is not as drastic as the diagonal case. 
As the eigenmode analysis trims modes, it excludes information and the same input data 
produces a different statistical representation.  In light of this effect, a $1\sigma$ 
difference becomes a statistical difference of analysis rather than a significant systematic effect. 

In summary, we find lower galaxy-mass bias parameters with larger bin-widths, a potential 
artificial bias on the galaxy-mass parameters due to over-smoothing.  Since we gain very 
little additional constraining power with the $f=0.25$ bin-width, we argue the $f=0.10$ 
bin-width represents the more conservative choice.  Although the $f=0.10$ scheme represents 
smaller bins, they are still quite large and adequately resolve structure in the covariance.

\end{appendix}

\def\baselinestretch{1}
\bibliography{refs}

\end{document}